\documentclass[aps,prb,twocolumn,showpacs]{revtex4}

\usepackage{boxedminipage}  
\usepackage{lscape}     
\usepackage{float}      
\usepackage[ansinew]{inputenc}  
\usepackage{epsfig}
\usepackage{subfigure}
\usepackage{graphicx}
\usepackage{amsmath}
\usepackage{amssymb}

\renewcommand{\d}[2]{\frac{#1}{#2}}
\newcommand{\pd}{\partial}

\begin{document}

\hyphenation{Brillouin}

\title{Transport in the metallic regime of Mn doped III-V Semiconductors}

\author{Louis-Fran\c cois Arsenault}
\email{lfarsena@physique.usherbrooke.ca} \altaffiliation{Department
of physics, University of Sherbrooke}
\author{B. Movaghar}
\altaffiliation{Department of Electrical and Computer Engineering,
Northwestern University, Evanston, IL, USA}
\author{P. Desjardins}
\author{A. Yelon}

\affiliation{D\'epartement de G\'enie Physique and Regroupement
Qu\'eb\'ecois sur les Mat\'eriaux de Pointe (RQMP)\\ \'Ecole
Polytechnique de Montr\'eal, C.P. 6079, Succursale ``Centre-Ville'',
Montr\'eal (Qu\'ebec), H3C 3A7, Canada}

\date{\today}

\begin{abstract}
The standard model of Mn doping in GaAs is subjected to a coherent
potential approximation (CPA) treatment. Transport coefficients are
evaluated within the linear response Kubo formalism. Both normal
(NHE) and anomalous contributions (AHE) to the Hall effect are
examined. We use a simple model density of states to describe the
undoped valence band. The CPA bandstructure evolves into a spin
split band caused by the $p-d$ exchange scattering with Mn dopants.
This gives rise to a strong magnetoresistance, which decreases
sharply with temperature. The temperature ($T$) dependence of the
resistance is due to spin disorder scattering (increasing with $T$),
CPA bandstructure renormalization and charged impurity scattering
(decreasing with $T$). The calculated transport coefficients are
discussed in relation to experiment, with a view of assessing the
overall trends and deciding whether the model describes the right
physics. This does indeed appear to be case, bearing in mind that
the hopping limit needs to be treated separately, as it cannot be
described within the band CPA.
\end{abstract}


\maketitle

\section{Introduction\label{sec:intro}}
Magnets, which can be made using semiconductors by doping with
magnetic impurities, are potentially very important new materials
because they are expected to keep some of the useful properties of
the host system. One important question is how much of the "doped
semiconductor" bandstructure does the system still have? Is it a
completely new alloy or does the system still behave like a doped
semiconductor, which can, for example, be described using Kane
theory, as taught in standard textbooks \cite{Chuang}. Many of these
questions have been seriously investigated in extensive recent
reviews \cite{Jungwirth1,Sinova1,Ohno1,Ohno2}. The objective of this
paper is to present a simple and mostly analytically tractable
description of the magnetism and transport properties of GaMnAs. We
need a theory which captures the essential physics, can be
parameterized and applied to magnetic nanostructures without having
to carry out ab-initio calculations each time some parameter has
changed. Previous theories of GaMnAs and related materials have been
either computational or semi-analytical. But the treatment has often
been split into three parts: magnetism, resistivity, and
magnetotransport have been treated separately. Here we show that the
magnetism and transport can be treated on the same footing using a
one-band model. The one band model is of course not enough to
explain the optical properties. But we show that magnetism, spin,
and charged impurity scattering can be formulated in the same
framework, using the coherent potential approximation (CPA) and a
generalization thereof. In this paper we focus, however, only on the
magnetism and spin scattering aspects and leave the computation of
the charged impurity scattering and the full quantitative comparison
to a later paper. One of the most difficult and controversial
aspects for all conducting magnets is the origin of the Anomalous
Hall effect (AHE). This is why we have devoted a section to
reminding the reader of the basic definitions and results. The full
history of the AHE is given
in Ref.~\onlinecite{Sinova1}.\\
\\
We focus on GaMnAs as a much-studied prototype. It is now generally
accepted that the observed ferromagnetic order of the localized
spins in Mn-doped GaAs is due to Mn impurities acting as acceptor
sites, which generate holes. These are antiferromagnetically coupled
to the local 5/2 Mn spins, lowering their energy when they move in a
sea of aligned moments. It is the configuration of lowest energy
because spin-down band electrons move in attractive potentials of
spin-up Mn and vice versa. Given that the spins are randomly
distributed in space, the magnetic state is also the state of
maximum "possible order". Thus, free and localized spins are
intimately coupled. These materials exhibit strong negative
magnetoresistance because aligning the spins reduces the disorder,
and this lowers the resistance. The degree to which the
magnetization of the spins affects the scattering process is
dependent upon the degree to which the magnetic spin scattering is
rate-determining for resistance. We shall, in this paper, only
include the spin scattering process in order to first gain an
intuitive understanding of the processes involved.\\
\\
The "hole mediated" magnetism point of view is not shared by
everyone. Mahdavian and Zunger \cite{Mahdavian_Zunger} argue that
the mobile hole induced magnetism cannot explain the magnetism in
high band gap, strongly-bound, hole materials such as Mn doped GaN.
Their model, based on first principle supercell calculations,
predicts that the Mn induced hole takes on a more $d$-like character
as the host band gap increases, making the magnetism a "$d-p$"
coupling property.\\
\\
The material of this paper is structured as follows: we first
present a short review of the phenomenology of the Hall effect in
magnets. The Hamiltonian which describes the properties of Mn doped
semiconductors is then introduced. We then discuss, in more detail,
how to formulate transport in magnetically doped semiconductors.
Following that, we recall how one can calculate the self-consistent
CPA self-energy caused by the spin dependent term in the generally
accepted Hamiltonian, for a one band system. Once the self-energy is
known, we compute the longitudinal and transverse conductivities
using the Kubo formulae. When the Fermi level is just above the
mobility edge of the hole band, the Kubo formula is still valid and
describes the strong-scattering random phase limit. The localized
hopping regime is treated in another paper \cite{Arsenault1}. The
CPA equations only need the density of states of the active free
hole band as input. We will therefore consider the semicircular
Hubbard band which reproduces the correct free hole band edges and
is mathematically convenient for illustrating the effects of band
induced magnetism.
\section{The Hall effect\label{sec:Halleff}}
The experimentally measured Hall coefficient $R_H$ is often written
as
\begin{equation}\label{eq:RHall_1}
\begin{split}
R_H &= R_N + R_A\\
R_A &= \left(a\rho_{xx}+b\rho_{xx}^2\right)M/B_z,
\end{split}
\end{equation}
where $a$,$b$ are constants and $\rho_{xx}$ is the resistivity. The
first term $R_N$ is the normal Hall coefficient and scales with
resistivity in the usual way and the second $R_A$ is the anomalous
term, which in general can have two components, one linear and the
other quadratic with resistivity\cite{Jungwirth1} and is
proportional to the magnetization $M$. The general relation for the
Hall coefficient is
\begin{equation}\label{RHall_rigor}
R_H =
\d{\text{Re}\left\{\sigma_{xy}\left(B_z\right)\right\}}{B_z\left[
\text{Re}\left\{\sigma_{xx}\left(B_z\right)\right\} \right]^2},
\end{equation}
where $\sigma_{xy}$ and $\sigma_{xx}$ denote the transverse and
normal conductivity respectively and $B_z$ is the magnetic field.\\
\\
Karplus and Luttinger\cite{Karplus_Luttinger} pointed out that the
$B$-field involves the magnetic moment of the material via the
internal magnetization $M$
\begin{equation}\label{mag_field}
B_z = \mu_0\left[ H_z^{ext} + (1-N)M \right] \equiv B_z^{ext} +
\mu_0(1-N)M,
\end{equation}
where $N$ is the demagnetizing factor. Thus the magnetization term
is implicit in the normal contribution as a shift in the magnetic
field. However this form is not normally sufficient to explain the
much larger magnetization contribution observed in
ferromagnets\cite{Karplus_Luttinger}. Throughout we shall, for
simplicity, suppose a thin film with perpendicular-to-plane magnetic
field and thus take $N = 1$, unless otherwise mentioned.
\section{Microscopic description}
\subsection{Hamiltonian}
In this section we formulate the basic model. The Hamiltonian for
this Mn doped "alloy", which most workers in the field accept as the
correct description \cite{Jungwirth1}, is given by
\begin{equation}\label{Htot}
H_{tot} = H_{p} + H_{Loc} + H_d + H_{pd} + H_{p-B} + H_{d-B} +
H_{so}.
\end{equation}
In Eq.~\eqref{Htot}, the first term describes the free band holes.
In the local Wannier or tight binding representation, it is
\begin{equation}\label{H_p} H_p =
\sum_{m,n,s}t_{mn}c_{ms}^{\dagger}c_{ns},
\end{equation}
with $t_{mn}$ denoting the overlap or jump terms from sites $m$ to
$n$, the $c_{ns}^{\dagger}$, $c_{ns}$ are creation and annihilation
operators for a carrier of spin $s$ at site $n$, and where the
indices include multiband transfers, if any. In the presence of a
magnetic field, the overlap has a Peierls phase, so that we write it
as (assuming thin film geometry, no internal field effect for a
field perpendicular to plane)
\begin{equation}\label{t_pei}
t_{mn} =
t^0_{mn}\text{e}^{\d{-ie}{2\hbar}\textbf{B}^{ext}\cdot(\textbf{R}_n\text{x}\textbf{R}_m)},
\end{equation}
where $ext$ stands for external, $e$ is the electronic charge and
$\textbf{R}_n$ the position vector of site $n$. The second term in
Eq.~\eqref{Htot} describes the diagonal (atomic orbital) energy that
the valence band p-hole experiences when it is sitting on an
impurity site which may or may not be magnetic
\begin{equation} H_{Loc} =
\sum_{m,s}E_{m,\alpha}c_{ms}^{\dagger}c_{ms}, \end{equation} where
$\alpha$ stands for magnetic ($M$) or non-magnetic ($NM$).\\
\\
The third term is the direct exchange coupling between the Mn
d-localized spins, which we neglect here because when mobile
carriers are present, they mediate the exchange coupling and in the
case of a low concentration, the Mn spins should be, on average, far
away from each other. The fourth term is the antiferromagnetic spin
exchange coupling between the valence p-hole and the local d-Mn and
ultimately the reason for magnetism in these materials
\cite{Dietl1,Okabayashi,Szczytko}
\begin{equation}\label{double_echange}
H_{pd} = \d{J_{pd}}{2}\sum_{m \in
\{N_S\}}\sum_{s,s'}c_{ms}^{\dagger}\boldsymbol{\sigma}_{ss'}\cdot\textbf{S}_mc_{ms'},
\end{equation}
where $\boldsymbol{\sigma}_{ss'}$ is a vector containing the Pauli's
matrices i.e. $[\sigma^x, \sigma^y, \sigma^z]$. The indices, $s$ and
$s'$ indicate which terms of the 2x2 matrix we are considering.
Finally, $\textbf{S}_m$ is the Mn spin operator at site m. The fifth
and sixth terms are the Zeeman energies of the holes and Mn spins in
an external magnetic field along the z-axis, $B_z^{ext}$.
\begin{equation}\label{HpB}
H_{p-B} =
-\d{g^*}{2}\mu_B\sum_{m}\sum_{s}B_{z,m}^{ext}c_{ms}^{\dagger}\boldsymbol{\sigma}_{ss}^zc_{ms},
\end{equation}
\begin{equation}
H_{d-B} = -2\mu_B\sum_{m\in \{N_s \}}B_{z,m}^{ext}S^z_m,
\end{equation}
where $g^*$ is the effective g-factor. Finally we have the
spin-orbit coupling
\begin{equation}\label{Hso}
H_{so} = \sum_{m,n}\sum_{s,s'}\langle s,m|h_{so}|n,s'\rangle
c_{ms}^{\dagger}c_{ns'},
\end{equation}
where
\begin{equation}
h_{so} = \d{\hbar}{4m^2c^2}\boldsymbol{\sigma}\cdot\left(\nabla
V(\textbf{r})\times\textbf{p} \right)
\end{equation}
and where $V(\textbf{r})$ is the total potential acting at a point
$\textbf{r}$ and will include both normal crystal host sites and
impurity sites and $\textbf{p}$ is the momentum operator. Finally,
$c$ is the speed of light. In the tight binding representation,
disorder is normally in the diagonal energies. For the spin-orbit
term, disorder will enter the Hamiltonian through variations in the
site potential.
\subsection{Spin-orbit coupling}
It is generally believed that the spin-orbit coupling is the cause
of the anomalous Hall effect. This is one of the most fascinating
and universal observations made in conducting magnets. In this
section we briefly discuss the basic phenomenology of spin-orbit
coupling. It is useful to recall the basic premises.
\\\\
When the Bloch wavefunction ansatz is inserted into the Hamiltonian,
the spin-orbit coupling produces two new terms which enter the
Hamiltonian for the periodic part of the wavefunction
"$u_{n\textbf{k}}$" {$\Psi_{n\textbf{k}}(\textbf{r}) =
u_{n\textbf{k}}\text{e}^{i\textbf{k}\cdot\textbf{r}}$:
\begin{equation}
    h_{so} = \d{\hbar}{4m^2c^2}\Big[ \big( \nabla V(\textbf{r})\times\textbf{p} \big)\cdot\boldsymbol{\sigma} + \hbar\big( \nabla V(\textbf{r})\times\textbf{k} \big)\cdot\boldsymbol{\sigma}
    \Big].
\end{equation}
The first is the usual term and enters the band structure
calculation. The second is called the Rashba
term\cite{Bychkov_Rashba} and in a crystal is, in general, less
important than the first, because in the nucleus, momenta are larger
than the lattice momenta $\textbf{k}$.\\
\\
The "Rashba term" is not normally treated in the computation of
$u_{n\textbf{k}}$. But in a lattice, it may be enhanced. In the Kane
model, Chazalviel\cite{Chazalviel} and De Andrada e Silva \emph{et
al}.\cite{DeAndrada} have shown how to handle the spin-orbit terms
in a lattice and the Rashba term in the presence of a triangular
potential produced by a gate in an inversion layer. In particular,
Ref.~\onlinecite{DeAndrada} have shown how to renormalize the
coefficient of the Rashba term and that this term acts when
inversion symmetry is broken in an external field or triangular
potential, for example. The origin of the enhancement of both these
spin-orbit effects can perhaps be understood as similar to the
origin of the change of the effective mass in the lattice. The
spin-orbit energy can be enhanced by the presence of the lattice,
but the enhancement can be very different from case to case, and
needs to be examined in each case separately. For example, naively
speaking, one may consider the spin-orbit force acting on a particle
moving in a slowly varying potential of longer range than the
lattice spacing as a spin current of an effective mass electron. One
can think of the scattered particle as having an effective mass
$m^*$ and generating a spin-orbit interaction which scales as
$\d{\hbar}{4(m^*c)^2}$ instead of $\d{\hbar}{4(mc)^2}$. When the
particle is scattered from an atomic size impurity in a lattice, it
is forced to come back many times and spends a longer time on the
impurity than in free space. Leaving aside these intuitive pictures,
one can, in any case, make the analysis quantitative using the Kane
Hamiltonian, which takes the lattice modulation into account in a
kind of renormalized perturbation theory and gives explicit results
for the $u_{n\textbf{k}}$ part of the wavefunction\cite{Chuang}. The
Kane Hamiltonian can then be used to treat scattering from an
impurity. Thus, for example, the matrix elements of the position
operator $\langle\textbf{k}\big|\textbf{r}|\textbf{k}'\rangle$ are
very different for plane wave states and for the Kane
solutions\cite{Chazalviel,Karplus_Luttinger}.\\
\\
In tight-binding (TB), the spin-orbit matrix elements are calculated
using atomic orbitals, with two terms: the intra- and the
inter-atomic contributions. The magnitude of the intra-atomic term
is just what it would be for the corresponding orbitals on the atoms
in question. The inter atomic term is normally small and not
included in tight-binding band structure calculations. When an
electron jumps from site to site, it experiences a net magnetic
field produced via the cross product of its velocity and the
electric field gradient of the neighboring ions. This field then
couples to its spin.  Since it is related to the intersite momentum
or velocity, it depends on the transfer rate $t_{mn}$  where the
indices $m$,$n$ include a band index {$m = m,\gamma$}. Remember that
the velocity ($x$-direction) operator is given by
\begin{equation}\label{vx}
    v_x = \d{i}{\hbar}\sum_{m,n,s}\left( \textbf{R}_{m,n}
    \right)_xt_{mn}c_{ms}^{\dagger}c_{ns},
\end{equation}
where $\textbf{R}_{m}$ is the position vector of site $m$ and
$\textbf{R}_{m,n} = \textbf{R}_{m} - \textbf{R}_{n}$. An enhancement
of this spin-orbit energy, if any, has to be calculated by taking
the expectation value of the spin-orbit term using the calculated
Bloch states as one would for the dispersion relation
$\varepsilon_{ks}$ and for the effective masses. We shall examine
the spin orbit coupling in more detail below.\\
\\
\section{The Kubo transport equations}
In order to compute the transport properties of the magnetically
doped semiconductors in the linear response regime, we need to
introduce the Kubo formulae. In this section we show how to compute
the transport coefficients.
\subsection{General considerations}
The conductivity in linear response to an electric field is usually
written as\cite{Movaghar_Cochrane1,Movaghar_Cochrane2,Roth}
\begin{equation}\label{kubo} \sigma_{\mu\nu} =
\d{i\hbar e^2}{\Omega}\lim_{\delta \rightarrow
0}\sum_{\alpha,\beta}\d{\langle\alpha|v_{\mu}|\beta\rangle\langle\beta|v_{\nu}|\alpha\rangle}{\varepsilon_{\alpha}
- \varepsilon_{\beta} + \hbar\omega +
i\delta}\d{f(\varepsilon_{\alpha}) -
f(\varepsilon_{\beta})}{\varepsilon_{\beta} - \varepsilon_{\alpha}},
\end{equation}
where the $v_{\mu}$ are the velocity operators in the respective
direction $\mu$. For a given Hamiltonian they are derived from the
Heisenberg relation $i\hbar v_{\mu} = [x_{\mu},H]$ with $x_{\mu}$
the position operator. The $|\alpha\rangle$ and
$\varepsilon_{\alpha}$ are the exact wavefunctions and energy
levels. The wavefunctions include any disorder and magnetic and
spin-orbit couplings. $f(\varepsilon)$ is the Fermi function and
$\omega$ is the frequency of the applied electric field; $e$ is the
electronic charge. This formula can now be written in the particular
representation selected, i.g. tight-binding (TB) or Bloch states.
Within the bandstructure approach for semiconductors, one would
substitute the 8-band $\textbf{k}\cdot\textbf{p}$ wavefunctions and
energies into Eq.~\eqref{kubo}, include spin splitting through a
Weiss field and spin-orbit coupling, and treat disorder as a
lifetime contribution in the energy levels. In this picture the
doped semiconductor retains the pure band structure features, apart
from a complex lifetime shift. We shall use the one band CPA for the
coupling of the holes to the magnetic impurities. For the
calculation of the ordinary transport coefficients we may neglect
the spin-orbit coupling. We obtain, for the longitudinal
conductivity of the one band TB model for $\omega = 0$ and per spin
$s$
\begin{equation}\label{conduc}
\begin{split}
\langle\sigma^{s}_{xx}\rangle &= \\ &\d{e^2\hbar}{\pi \Omega}\int
dE\left( -\d{\pd f(E)}{\pd E} \right)\int d\varepsilon
X(\varepsilon)\bigg[ \text{Im}\{ \langle G_{s}(E,\varepsilon)\rangle
\} \bigg]^2,
\end{split}
\end{equation}
where
\begin{equation}\label{Green_def}
\langle G_{s}(E,\varepsilon)\rangle = \d{1}{E-\varepsilon-\Sigma_s},
\end{equation}
\begin{equation}
\Sigma_s = \Sigma_{s[R]} - i\Sigma_{s[I]},
\end{equation}
and
\begin{equation}
X(\varepsilon) =
\d{1}{N}\sum_kv_x^2\delta(\varepsilon-\varepsilon_k).
\end{equation}
$\Sigma_s(E)$ is the CPA self-energy for spin state $s$, to be
calculated self-consistently using the CPA condition and where the
velocity and effective masses of the free electron band are given by
\begin{equation}
v_{\mu} = \d{1}{\hbar}\d{\pd\varepsilon_k}{\pd k_{\mu}},
\end{equation}
\begin{equation}
M_{\mu\nu}^{-1} = \d{1}{\hbar^2}\d{\pd^2\varepsilon_k}{\pd
k_{\mu}\pd k_{\nu}},
\end{equation}
where $M_{\mu\nu}$ is the effective mass tensor. The CPA self-energy
contains the information on the net spin splitting produced by the
magnetic impurities and the scattering time. It defines the
effective medium which is to be discussed below.\\
\\
The Hall conductivity, given by the antisymmetric part of the
transverse conductivity\cite{Matsubara,Ballentine}, is somewhat more
complicated because the magnetic field has to be dealt with first.
To first order in magnetic field, and with one TB band we find for
the antisymmetric part (referred to by the index $a$) for $\omega =
0$\cite{Roth,Movaghar_Cochrane1,Matsubara}:
\begin{equation}\label{conduc_Hall}
\begin{split}
&\langle \sigma^{a}_{xy}\rangle =\\ &\d{2e^3\hbar^2B}{3\pi
\Omega}\sum_s\int dE\left( -\d{\pd f(E)}{\pd E}\right)\int
d\varepsilon Y(\varepsilon)\bigg[\text{Im}\{\langle
G_{s}(E,\varepsilon)\rangle\bigg]^3,
\end{split}
\end{equation}
where
\begin{equation}
Y(\varepsilon) \equiv \d{1}{N}\sum_k\left[ \d{v_x^2}{M_{yy}} +
\d{v_y^2}{M_{xx}} - \d{2v_xv_y}{M_{xy}} \right]\delta(\varepsilon -
\varepsilon_k).
\end{equation}
In Eq.~\eqref{conduc_Hall}, the dispersion relation $\varepsilon_k$
can be any one band structure (any choice of $t_{mn}^0$) where
$\varepsilon_k$ is defined through $t_{mn}^0 =
\d{1}{N}\sum_k\text{e}^{i\textbf{k}\cdot\textbf{R}_{mn}}\varepsilon_k$.
As was shown by Matsubara and Kaneyoshi\cite{Matsubara}, for weak
magnetic field, the Peierls phase can be factored out of the Green
function and this is why we only need the Green function
(Eq.~\eqref{Green_def}) that does not include the orbital effect of
the field. The effect of the different phases (from the $t_{mn}$ and
from the Green functions), after an expansion to first order in the
field, are reflected in the topology dependent term
$Y(\varepsilon)$. Furthermore, Eq.~\eqref{conduc_Hall} does not
contain the spin-orbit scattering contribution. It can be included
as an extrinsic effect. In the spirit of the skew scattering Born
approximation\cite{Smit,Ballentine,Engel} and in the weak scattering
limit, we can, at the end of the calculation, introduce an extra
internal magnetic field so that the total field entering
Eq.~\eqref{conduc_Hall} is
\begin{equation}\label{tot_mag_field}
B = B_z + \d{m^*}{e}\left\langle \d{1}{\tau_s} \right\rangle\langle
\sigma_z \rangle
\end{equation}
where $B_z$ includes the internal magnetization. However, from
Eq.~\eqref{mag_field}, for a field perpendicular to plane, $N = 1$.
This result, Eq.~\eqref{tot_mag_field}, is written in the notation
of Ballentine\cite{Ballentine}. The effective spin-orbit field
depends on the thermally averaged spin polarization $\langle
\sigma_z \rangle$, with its sign to be discussed in
Section~\ref{subsec:AHE}, and $\left\langle \d{1}{\tau_s}
\right\rangle$ is the so-called skew scattering rate. The second
term of Eq.~\eqref{tot_mag_field} involves, in addition to the sign
of the carrier, the sign of
the spin orientation.\\
\\
Engel \emph{et al}.\cite{Engel} claim that the impurity spin-orbit
coupling in the lattice environment can be "6 orders or magnitude"
larger than in vacuum. The large enhancement can mean that, in some
cases, the skew scattering dominates where there should exist a
substantial intrinsic contribution as well\cite{Engel,Kato}. This is
why developing a way to calculate the extrinsic contributions to the
anomalous Hall effect, as we do here, is important and useful.
\subsection{The Anomalous Hall Effect (AHE)}\label{subsec:AHE}
Let us now show how one can derive such an anomalous Hall effect, in
our formalism (TB approximation), from the existence of a spin-orbit
coupling and how one can arrive at the concept of an effective
magnetic field, as given in
Eq.~\eqref{tot_mag_field}.\\
\\
In tight-binding, the spin-orbit coupling is given by
Eq.~\eqref{Hso} and $V(\textbf{r})$ is the total potential
experienced by the charge at point $\textbf{r}$, with $\textbf{p}$
the momentum operator. The spin-orbit coupling can be separated into
an "intra" and an "inter" atomic contribution by splitting the
momentum operator as $\textbf{p} = \textbf{p}_a +
\textbf{p}_{inter}$. For spherically symmetric potentials the two
terms are given by
\begin{equation}\label{Vso}
V_{so} =
\sum_i\lambda_i(\textbf{r})\textbf{l}_i\cdot\boldsymbol{\sigma} +
\sum_n\Big\{
(\textbf{r}-\textbf{R}_n)\lambda_n(\textbf{r})\times\textbf{p}_{inter}
\Big\}\cdot\boldsymbol{\sigma},
\end{equation}
\begin{equation}
\lambda_n(\textbf{r}) = \d{\hbar}{4m^2c^2}\Bigg|\d{\nabla
V_n(\textbf{r}-\textbf{R}_n)}{\textbf{r}-\textbf{R}_n}\Bigg|,
\end{equation}
where $\lambda_n(\textbf{r})$ is related to the spin-orbit coupling
strength. Here $\textbf{p}_{inter}$ is the usual zero-field
inter-atomic momentum operator, $\textbf{l}_i$ is the atomic orbital
operator and is often quenched so that its expectation value is
zero, but nondiagonal same site inter-orbital matrix elements can be
very important.\\
\\
The second contribution in Eq.~\eqref{Vso} is due to the
inter-atomic motion, and must be evaluated using $\textbf{p}_{inter}
= m\textbf{v}$ where $\textbf{v}$ is given by Eq.~\eqref{vx} and
involves the intersite transfer energy $t$ and position operator
\begin{equation}
\textbf{r} = \sum_{m,s}\textbf{R}_mc_{ms}^{\dagger}c_{ms}.
\end{equation}
Here we have a spin-orbit coupling only because the particle can
jump to another orbital. When substituting the site representation
in the second term of Eq.~\eqref{Vso}, and assuming spherical
symmetric potentials, either impurity with effective local charge
$eZ_n = eZ_{imp}$ or host $eZ_h$ we have
\begin{widetext}
\begin{equation}
\langle i|V_{so}^{(2)}|j\rangle = \d{\hbar}{4mc^2}\left\{
\left\langle
i\left|\sum_{n,l}\left[\d{eZ_n}{(4\pi\varepsilon_0)\left|
\textbf{r}-\textbf{R}_n
\right|^3}(\textbf{r}-\textbf{R}_n)\right|l\right\rangle\d{i}{\hbar}\times\textbf{R}_{lj}t_{lj}
\right]\right\}\cdot\boldsymbol{\sigma},
\end{equation}
\end{widetext}
where $i$ is the imaginary number while $\langle i|$ is the orbital
at site i. The analysis of this term is not trivial, and depends on
the lattice and potential distribution in question. If one has
decided that the extrinsic contributions are dominant, then one only
needs to sum around the impurities. In general, the simplest
approach is to take only the largest terms in the sum, i.e. the
diagonal terms $l = i$. In this case we get
\begin{widetext}
\begin{equation}\label{mat_el_so}
\begin{split}
\langle i|V_{so}^{(2)}|j\rangle &= i\langle i|\Phi_{\sigma}|j\rangle
t_{ij},\\
i\langle i|\Phi_{\sigma}|j\rangle &= \d{\hbar}{4mc^2}\left\{
\left\langle i\left|\sum_{n}\left[\d{eZ_n}{(4\pi\varepsilon_0)\left|
\textbf{r}-\textbf{R}_n
\right|^3}(\textbf{r}-\textbf{R}_n)\right|i\right\rangle\d{i}{\hbar}\times\textbf{R}_{ij}
\right]\right\}\cdot\boldsymbol{\sigma},
\end{split}
\end{equation}
\end{widetext}
It is convenient to treat this effect as a spin dependent phase in
the transfer, in analogy to the Peierls phase, to first order.
Indeed, to first order $\text{e}^x\approx 1+x$ so that we can write
$i\Phi_{ij} = \text{e}^{i\Phi_{ij}}-1$ and $H_p+H_{so}$ as
\begin{equation}
H_p+H_{so} = \sum_{i,j}\sum_{\sigma,\sigma
'}t_{ij}\langle\sigma|\text{e}^{i\Phi_{ij}}|\sigma '\rangle
c_{i\sigma}^{\dagger}c_{j\sigma '}.
\end{equation}
Then we also note that the $\textbf{R}_n = \textbf{R}_i$ and
$\textbf{R}_n = \textbf{R}_j$ terms in Eq.~\eqref{mat_el_so} are
zero after the vector product, leaving the gradient field
contributions due to the nearest neighbors as the largest of the
remaining terms in the $n$-sum. A similar result has been derived in
the context of the Rashba coupling by Damker et al.\cite{Damker}.\\
\\
With Eq.~\eqref{mat_el_so} we can rewrite the spin-orbit matrix
element as
\begin{equation}
h_{ij}^{so} = i\d{e}{2\hbar}t_{ij}\sum_{n,n\neq
i,j}(\textbf{R}_j\times\textbf{R}_{in})\cdot\textbf{B}_{n,so},
\end{equation}
where $\textbf{R}_{in} = \textbf{R}_{i} - \textbf{R}_{n}$ and where
we have introduced a spin-orbit magnetic field defined by
\begin{equation}\label{Bso}
\textbf{B}_{n,so}(i\rightarrow j) = \d{\hbar}{2mc^2}\left\langle
i\left|\d{Z_n}{4\pi\varepsilon_0|\textbf{R}_i-\textbf{R}_n|^3}
\right|i\right\rangle\boldsymbol{\sigma}.
\end{equation}\\
\\
In summary, let us write in this approximation, a new total phase
for Eq.~\eqref{t_pei}
\begin{equation}\label{phase}
\begin{split}
&\varphi_{ij} =\\
&\d{e}{2\hbar}\left(-\textbf{B}_z\cdot(\textbf{R}_j\times\textbf{R}_i)
+ \sum_{n,n\neq
i,j}(\textbf{R}_j\times\textbf{R}_{in})\cdot\textbf{B}_{n,so}\right).
\end{split}
\end{equation}
We thus see what the spin-orbit coupling does to the TB Hamiltonian.
It is, in terms of energy, a small effect. Its effect on the band
structure and magnetism will be neglected. In order to obtain
Eq.~\eqref{conduc_Hall} with $B$ defined as
Eq.~\eqref{tot_mag_field} one as to decouple the trace over the spin
polarization from the remaining expression and this automatically
makes $\langle\sigma_z\rangle$ the overall mobile spin polarization.
In reality, the skew-scatterring is due to carriers near the Fermi
level and therefore $\langle\sigma_z\rangle$ should be the average
weighted at the Fermi level. We will return to this point when we
attempt to fit Eq.~\eqref{conduc_Hall} and Eq.~\eqref{tot_mag_field}
to experiment in Section~\ref{subsec:Hall_effect}.
\subsection{Transport coefficients for a simple cubic band}
To perform a calculation, one needs to specify a lattice topology
($t_{mn}^0$). The simplest one is a simple cubic lattice with
nearest neighbor hopping. The dispersion relation is given, for a
d-dimensional system, by $\varepsilon_k = -2t\sum_{\alpha =
1}^d\cos(k_{\alpha}\text{a})$ where a is the lattice constant of the
simple cubic. With this particular form, we can
show\cite{Arsenault2} that the longitudinal and Hall conductivities
at zero frequency can be written, using Eqs.~\eqref{conduc} and
\eqref{conduc_Hall}, with no approximations, as
\begin{widetext}
\begin{equation}\label{conduc_d_fin}
\langle \sigma_{xx}\rangle = \d{e^2}{\text{d}\pi\hbar
\text{a}^{\text{d}-2}}\sum_s\left[\int \left( -\d{\pd f(E)}{\pd E}
\right)\int \bigg[ \text{Im}\Big\{ \langle
G_{s}(E,\varepsilon)\rangle \Big\}
\bigg]^2\int_{-\infty}^{\varepsilon}-zD_0(z)dzd\varepsilon
dE\right],
\end{equation}
\begin{equation}\label{conduc_d_hall_fin}
\langle \sigma^{a}_{xy}\rangle = \d{4e^3 B}{3\pi
\text{d}(\text{d}-1)\text{a}^{\text{d}-4}\hbar^2}\sum_s\left[\int
\left( -\d{\pd f(E)}{\pd E} \right)\int \bigg[ \text{Im}\Big\{
\langle G_{s}(E,\varepsilon)\rangle \Big\}
\bigg]^3\varepsilon\int_{-\infty}^{\varepsilon}-zD_0(z)dzd\varepsilon
dE\right],
\end{equation}
\end{widetext}
where $\langle G_{s}(E,\varepsilon)\rangle$ is given by
Eq.~\eqref{Green_def}, d is the number of dimensions, $a$ stands
once again for antisymmetric and $D_0(z)$ is the density of states
of the pure crystal. In 3D, $D_0(z)$ can be found knowing that, for
the above $\varepsilon_k$, for a d-dimensional cubic lattice, the
Green function is\cite{Economu}
\begin{equation}\label{G_SC}
    G_0(E) = \d{1}{2\pi^2t}\int_0^{\pi}d\phi x\textbf{K}(x),
\end{equation}
where $\textbf{K}(x)$ is the complete elliptic integral of the first
kind, with $x = \d{4t}{E + i\delta \mp 2t\cos(\phi)}$.\\
\\
Now we can return to the calculation of the new bands generated by
the hole-spin coupling and for this we use the CPA. Later, we use
Eq.~\eqref{tot_mag_field} to estimate the magnitude of the skew
scattering Hall effect.

\section{The CPA equations for a one-band model with coupling to local spins}
In this section we show how to compute the new energy bands in the
presence of a high concentration of dopants, magnetic and
nonmagnetic. The magnetic coupling between the Mn spin and charge
and the holes in the valence band creates a new valence
band-structure. This new Mn-induced band will be magnetic at low
enough temperatures. Since the Mn are randomly distributed, we
cannot use Bloch's theorem. One way forward is to use the powerful
self-consistent single site approximation known as the CPA. The CPA
self-energy\cite{Velicky} $\Sigma(E)$ is determined by the condition
that if at a single site, the effective medium is replaced by the
true medium, then the configurationally averaged $t$-matrix produced
by scattering from the difference between the true medium and
effective medium potentials must vanish\cite{Velicky}.\\
\\
In the present case we have localized Mn spins of magnitude $S =
5/2$ so that there are 6 possible states $S_z = {-5/2,…,5/2}$ and
thus 7 possible sites in the system including the non-magnetic ones.
We have for a "spin up" carrier, the normal non magnetic sites with
concentration $(1-x)$ with local site energy $E_{NM}$ and magnetic
field interaction $-\d{g^*}{2}\mu_BB_{z,m}^{ext}$, and the magnetic
sites with concentration $x$ with local site energy $E_M$ and
magnetic coupling $\d{J_{pd}}{2}S_m^z-\d{g^*}{2}\mu_BB_{z,m}^{ext}$.
But there is also the possibility of a spin flip of both the carrier
and the impurity with interaction $\d{J_{pd}}{2}S_m^+$. For a "spin
down" carrier we have $E_{NM}$ at a nonmagnetic site and magnetic
field interaction $\d{g^*}{2}\mu_BB_{z,m}^{ext}$. For a magnetic
site we still have $E_M$ as the local site energy but the magnetic
coupling is now $-\d{J_{pd}}{2}S_m^z+\d{g^*}{2}\mu_BB_{z,m}^{ext}$
and the spin-flip interaction is given by $\d{J_{pd}}{2}S_m^-$. The
concentration of spin carrying impurities is defined by $x$ and is
$\d{N_S}{N_L}$ where $N_S$ is the number of sites with an impurity
while $N_L$ is the total number of sites in the system. The CPA
conditions are given by\cite{Takahashi_Mitsui,Arsenault2}
\begin{equation}\label{cpa1}
\begin{split}
(1-x)\langle t_{m\uparrow\uparrow}^{NM}\rangle_{Th} + x\langle t_{m\uparrow\uparrow}^{M} \rangle_{Th} &= 0\\
(1-x)\langle t_{m\downarrow\downarrow}^{NM}\rangle_{Th} + x\langle
t_{m\downarrow\downarrow}^{M} \rangle _{Th}&= 0,
\end{split}
\end{equation}
and the thermal average of an operator is
\begin{equation}\label{val_moy_nom_exacte}
\begin{split}
    \langle \hat{O}_m(S^z)\rangle &= \d{\text{Tr}\left(\hat{O}_m(S^z)\text{e}^{-\beta H_{\textbf{S}}} \right)}{\text{Tr}\left(\text{e}^{-\beta H_{\textbf{S}}} \right)}\\
     &= \d{\sum_{S^z} O(S^z)\text{e}^{-\beta h_m(S^z)S^z}}{\sum_{S^z}\text{e}^{-\beta h_m(S^z)S^z}}\\
     &\equiv \sum_{S^z}O(S^z)P(S^z).
\end{split}
\end{equation}
In order to find the spin dependent $t$ matrix operator for the CPA,
it is useful to rewrite the Hamiltonian as a 2x2 matrix in spin
space. The $t_{\uparrow\uparrow}$ and $t_{\downarrow\downarrow}$
operators are the diagonals element of the $t$-matrix. The
expressions for the $t$-matrix are essentially the same as the ones
obtained previously by Takahashi and Mitsui\cite{Takahashi_Mitsui}.
The minor difference is that in the present case, we include the
magnetic field, so that the Zeeman term enters the $t$ operator. The
inclusion of the Zeeman term is straightforward: one has to add,
when $N = 1$, $\d{g^*}{2}\mu_BB_{z,m}^{ext}$ with the appropriate
sign in the expressions of Ref.~\onlinecite{Takahashi_Mitsui}. To
calculate the CPA $t$-matrices, we need the effective medium Green's
function which is no longer dependent on the site $m$.
\begin{equation}
G_s(E)\equiv\langle G_{mms}(E)\rangle =
\d{1}{N}\sum_{\textbf{k}}\d{1}{E-\varepsilon_{\textbf{k}s}-\Sigma_s(E)}.
\end{equation}
In the previous equation, $G_{ijs}(E)$ refers to the Green's
function of a carrier of spin $s$ between sites $i$ and $j$ for one
particular configuration of disorder. The average is over all
possible realization of disorder.\\
\\
The local density of states is
\begin{equation}\label{DOS_LOC}
    D_s(E) = -\d{1}{\pi}\text{Im}\Big\{ \langle G_{mms}(E)\rangle
    \Big\},
\end{equation}
and the hole spin concentration with spin $s$ at site $m$ can be
written
\begin{equation}\label{spin_conc}
    \langle\text{p}_{ms}\rangle = \int dEf(E)D_s(E).
\end{equation}
Assuming mean field theory for the energy entering the Boltzmann
factor, the probability that the local spin has a value $S^z$ is
\begin{equation}\label{Prob}
    P(S^z)=\d{\text{e}^{-\beta h_mS^z}}{\sum_{S^z}\text{e}^{-\beta
    h_mS^z}},
\end{equation}
where
\begin{equation}\label{hm}
    h_m = \d{J_{pd}}{2}\left[ \langle\text{p}_{m\uparrow}\rangle - \langle\text{p}_{m\downarrow}\rangle
    \right] - 2\mu_BB_{z,m}^{ext}.
\end{equation}
To compute the CPA self-energy $\Sigma_s(E)$, we need to specify the
lattice dispersion of the hole band. In effect, it suffices to
specify the bare pure lattice density of states ($D_0(E)$) since the
real and imaginary parts are simply related. The Fermi level is
determined by the condition that the total hole concentration p is
known and fixed relative to the total impurity concentration with
$x$ as maximum value. In general, there will be fewer holes than
dopants, but the number is not known and remains a fit parameter, so
we have
\begin{equation}\label{conc_CPA}
    \text{p} = \int dEf(E)[D_{\uparrow}(E) + D_{\downarrow}(E)].
\end{equation}
This now allows us to compute the CPA self-energy self-consistently
and then to determine the new density of states via
Eq.~\eqref{DOS_LOC} and the local spin polarization and mobile spin
concentration via Eq.~\eqref{val_moy_nom_exacte} and
Eq.~\eqref{spin_conc}. Thus, we can determine the magnetization and
the transport coefficients, the resistivity and the conductivity as
a function of $B$ via Eq.~\eqref{conduc_d_fin}, and the
magnetoresistance as the relative change with $B$. Finally, the
normal and anomalous Hall $R_H$ can be obtained via
Eq.~\eqref{conduc_d_hall_fin}.
\section{Applications of the CPA}\label{App_CPA}
For numerical calculations, to simplify the analysis and for proof
of principle, instead of the $D_0(E)$ found with Eq.~\eqref{G_SC},
we use the Hubbard function, given by
\begin{equation}\label{DOS_Hub}
    D_0^{3D}(E) = \d{2\Theta (W-|E|)}{\pi W^2}\sqrt{W^2-E^2},
\end{equation}
where $W$ is half the bandwidth and $\Theta$ is the Heaviside step
function. Eq.~\eqref{DOS_Hub} has the same band edge behavior as the
$D_0(E)$ calculated with Eq.~\eqref{G_SC}. With the simple form of
Eq.~\eqref{DOS_Hub}, the first integration in
Eq.~\eqref{conduc_d_fin} and Eq.~\eqref{conduc_d_hall_fin} is
analytically tractable.
\subsection{dc conduction and magnetoresistivity}
Here we calculate the dc conductivity
$\langle\sigma_{xx}\rangle$ from Eq.~\eqref{conduc_d_fin} and the
magnetoresistance. The calculations have been done assuming $N = 1$
(thin film). The calculations for the magnetoresistance have been
performed without
taking into account the spin-orbit interactions.\\
\\
Figure~\ref{fig:DOS_B0J04x053EM0p08_T} and
Fig.~\ref{fig:DOS_B0J05x053EM0p03_T} are plots of the CPA density of
states for two values of p and $J_{pd}$ assuming that the scattering
is solely due to the spin potential, that is when $E_M = E_{NM} =
0$.
\begin{figure}
    \begin{center}
        \subfigure[]{\label{fig:DOS_B0J04x053EM0p08_T-a}\includegraphics[scale=0.27]{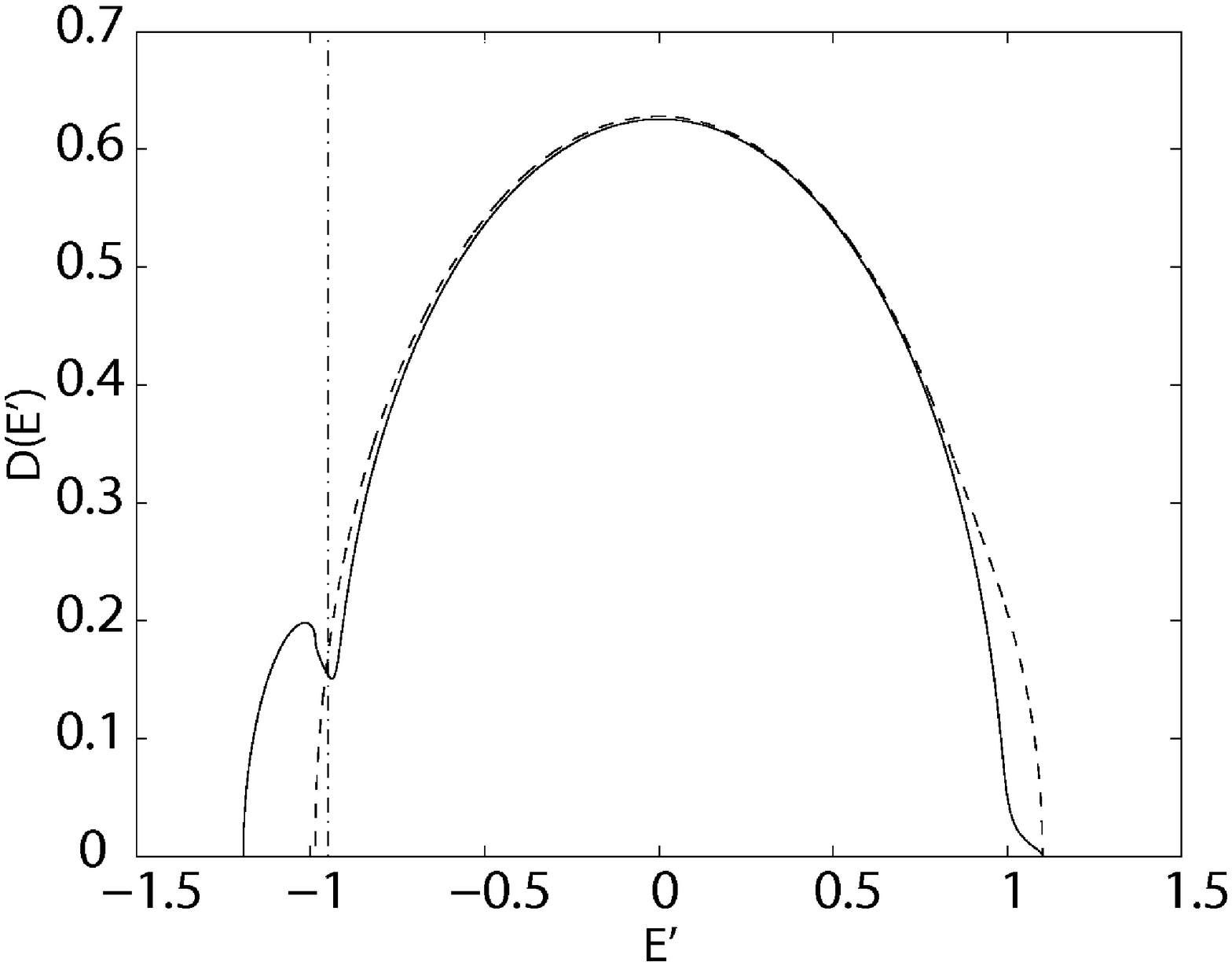}}
        \subfigure[]{\label{fig:DOS_B0J04x053EM0p08_T-b}\includegraphics[scale=0.27]{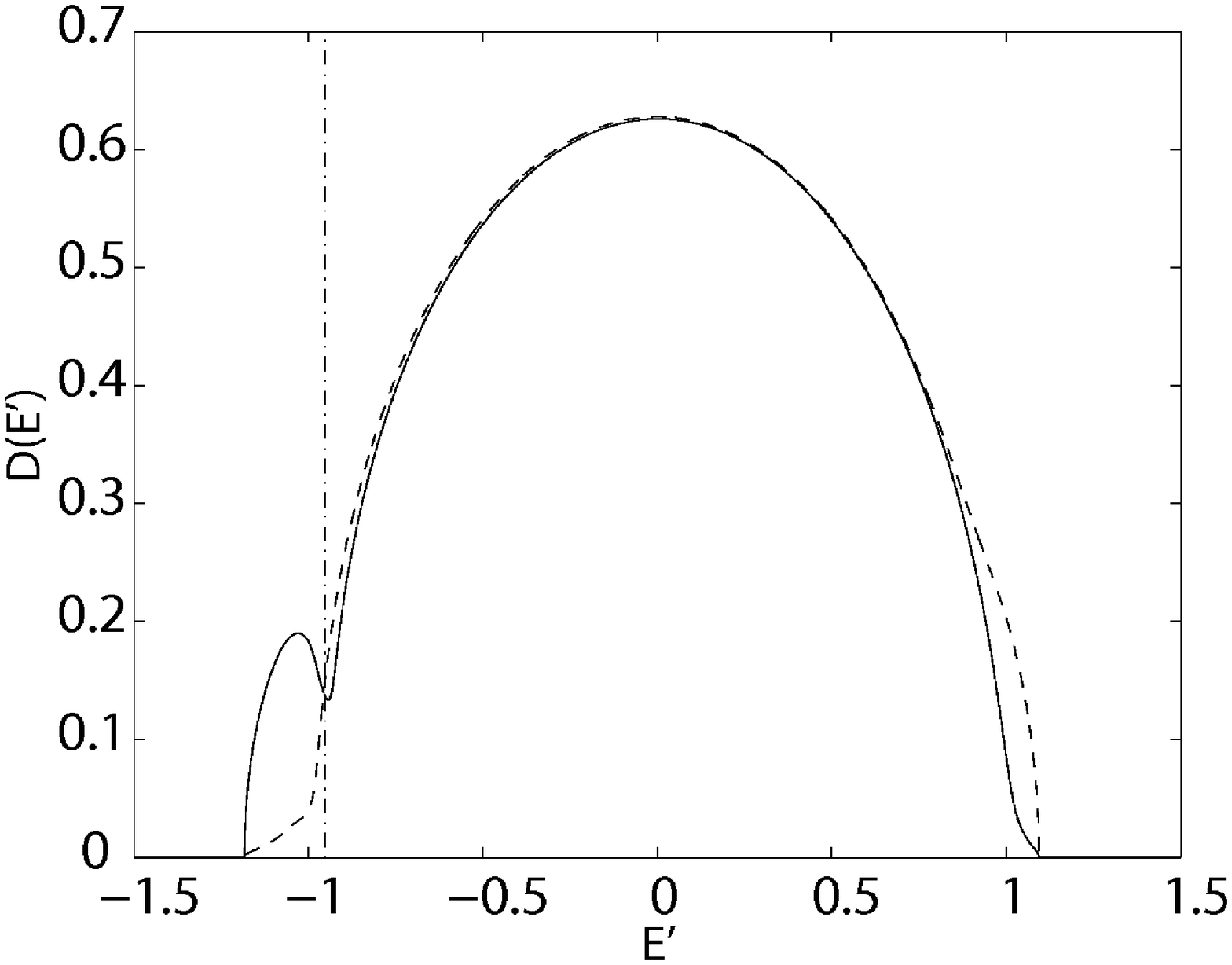}}
        \subfigure[]{\label{DOS_B0J04x053EM0p08_T-c}\includegraphics[scale=0.27]{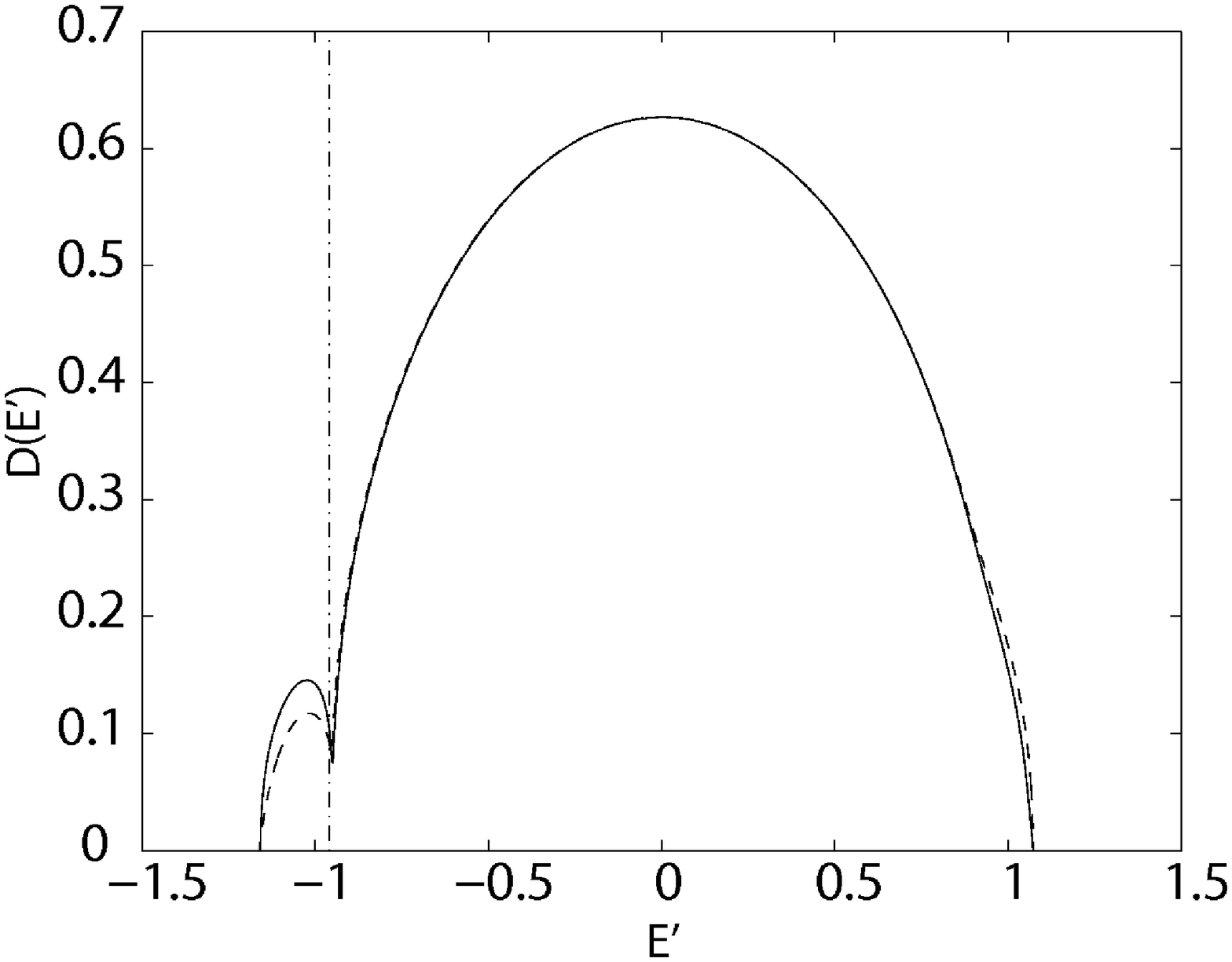}}
        \subfigure[]{\label{fig:DOS_B0J04x053EM0p08_T-d}\includegraphics[scale=0.27]{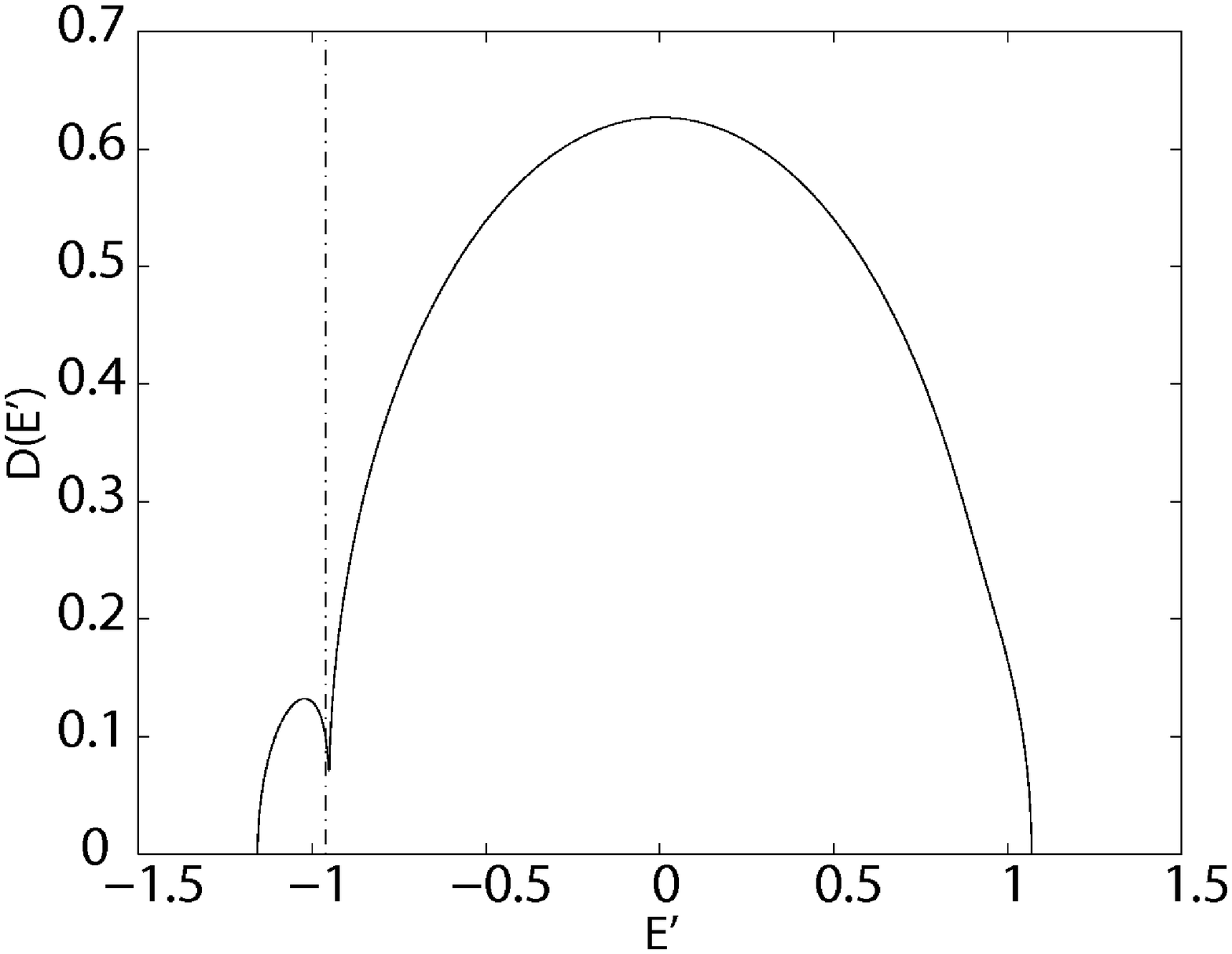}}
    \end{center}
      \caption{Density of states (DOS) for different temperatures, (a) $k_BT = 0$, (b) $k_BT = 5\text{x}10^{-3}W$,
      (c) $k_BT = 7.99\text{x}10^{-3}W$ and (d) $k_BT = 8.1\text{x}10^{-3}W$, when $x = 0.053$, $E_{NM} = E_M = 0$,
       $\mu_B B = 0$, $J_{pd} = 0.4W$  and $\text{p} = 0.8x$.  The full line curve is the DOS for the spin down carrier, the dashed one is the DOS for the spin up and the vertical line shows the Fermi level. One can see that the density of states is, relative to a nonmagnetic system, a strong function of temperature.}
  \label{fig:DOS_B0J04x053EM0p08_T}
\end{figure}
\begin{figure}
    \begin{center}
        \subfigure[]{\label{fig:DOS_B0J05x053EM0p03_T-a}\includegraphics[scale=0.27]{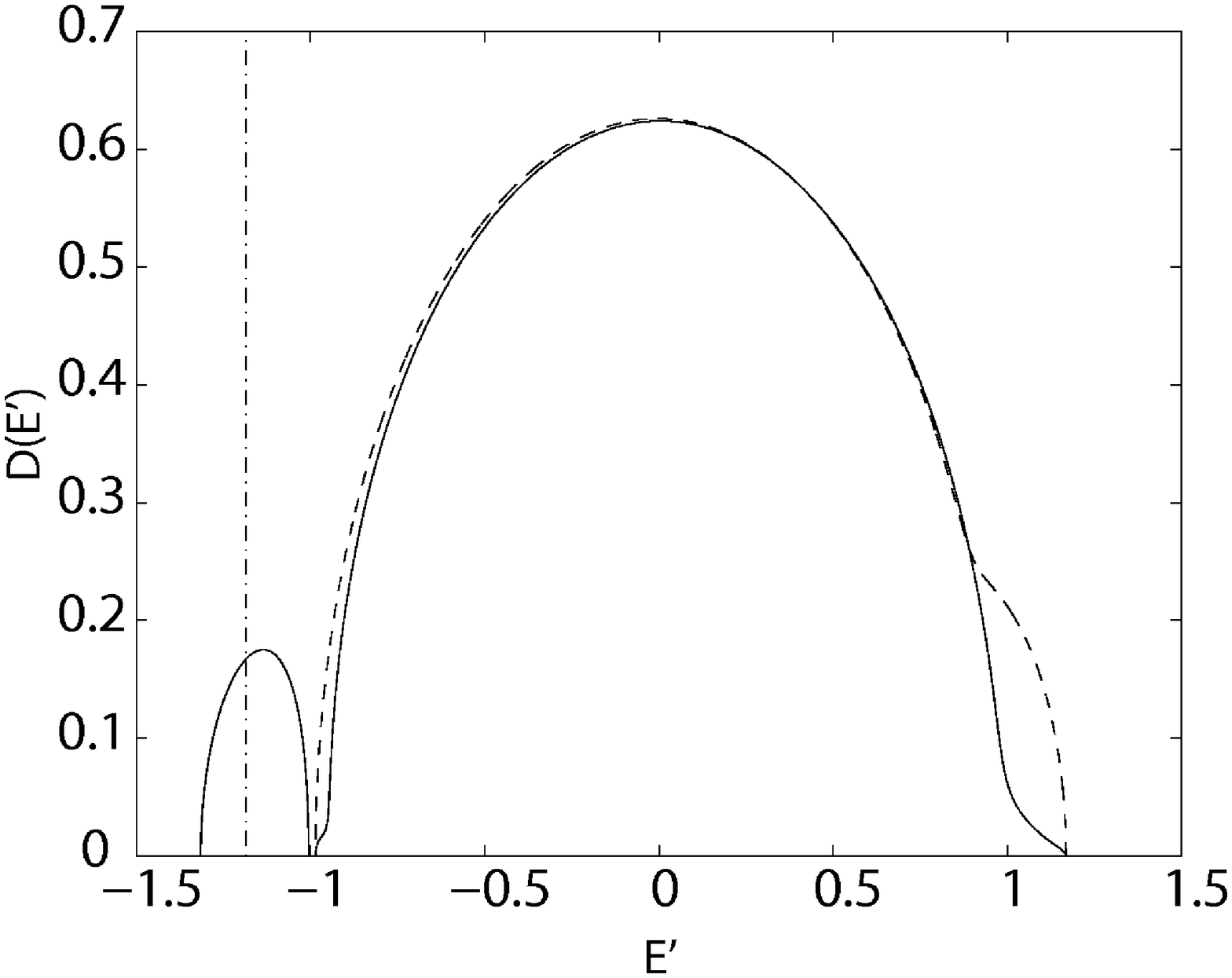}}
        \subfigure[]{\label{fig:DOS_B0J05x053EM0p03_T-b}\includegraphics[scale=0.27]{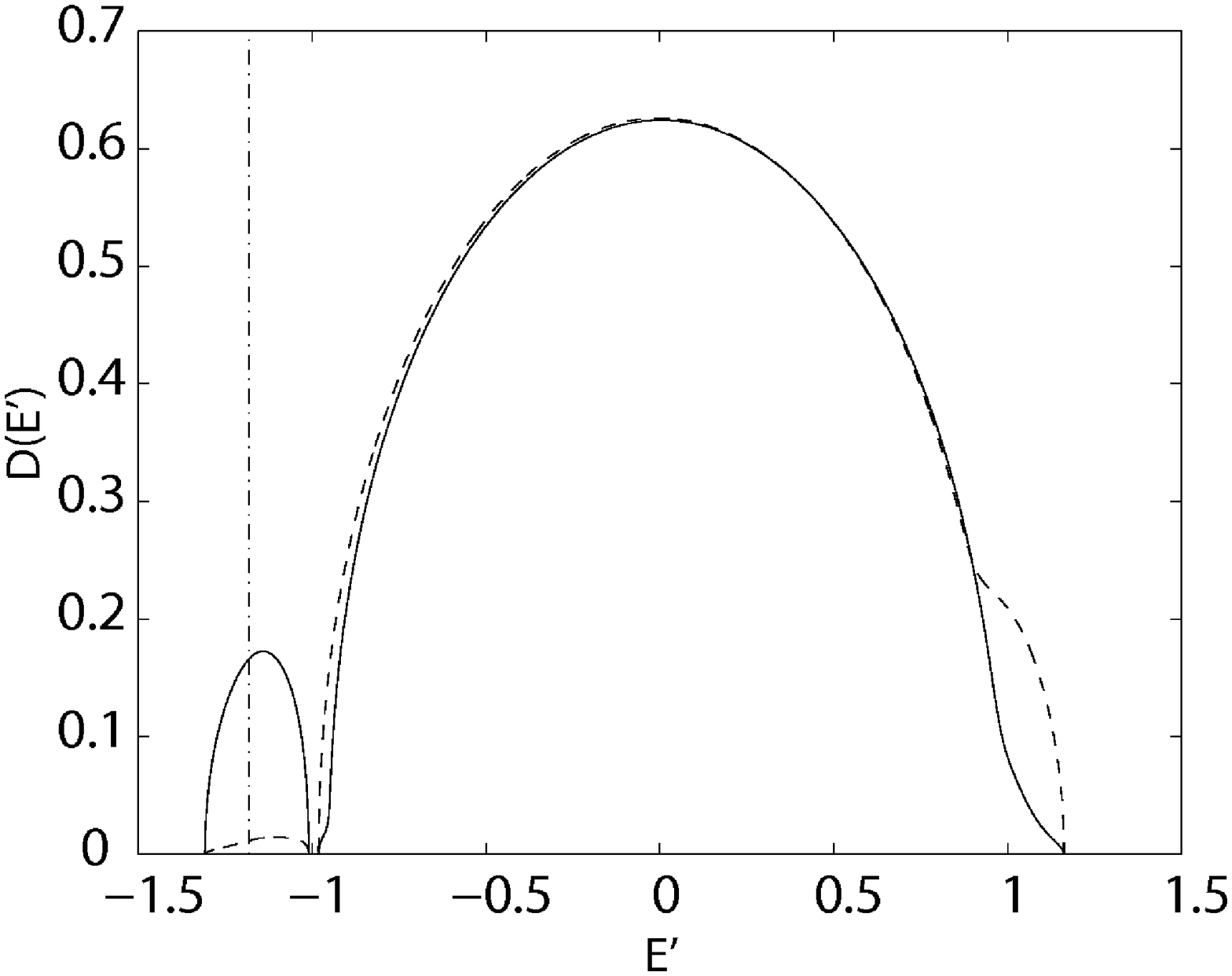}}
        \subfigure[]{\label{DOS_B0J05x053EM0p03_T-c}\includegraphics[scale=0.27]{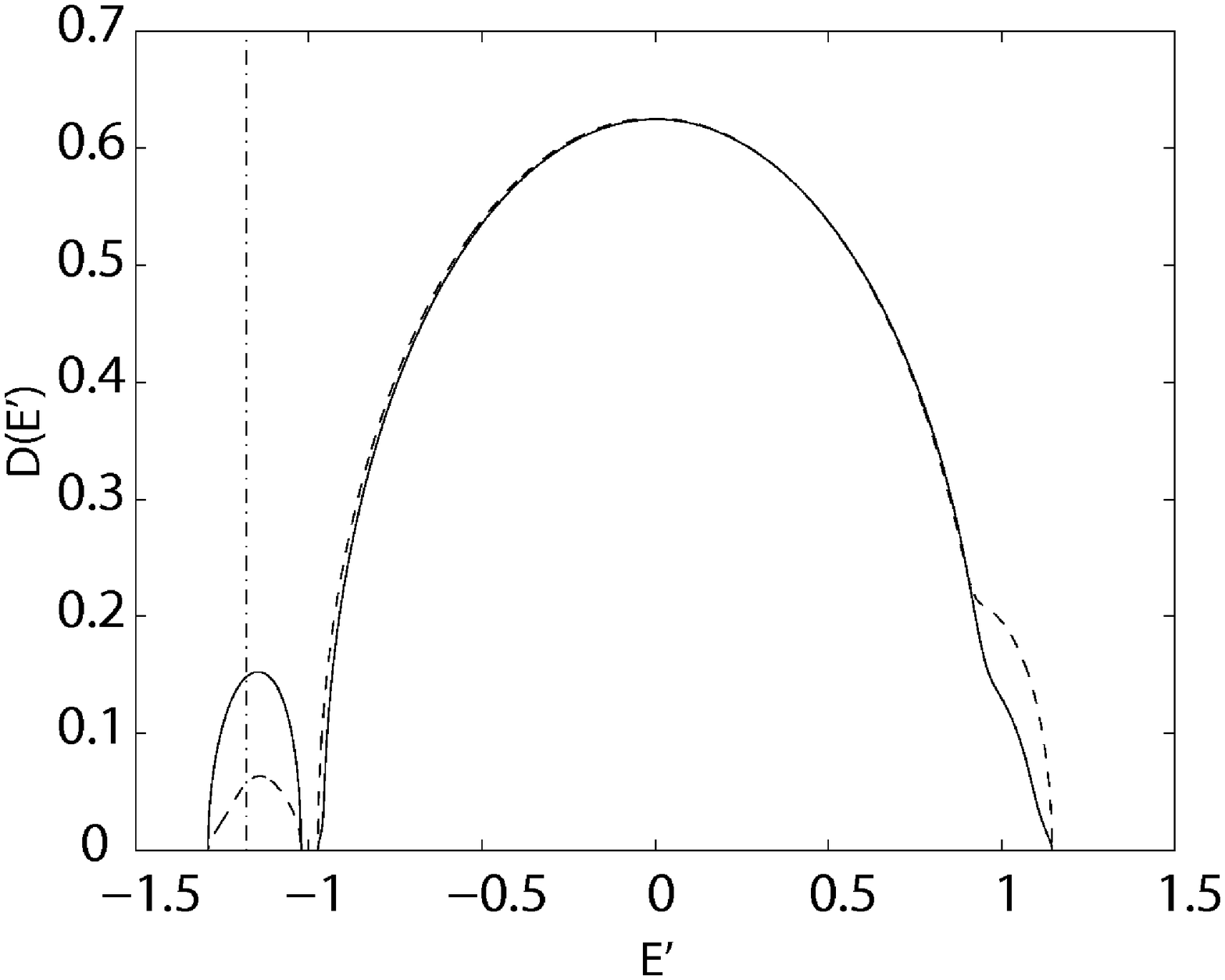}}
        \subfigure[]{\label{fig:DOS_B0J05x053EM0p03_T-d}\includegraphics[scale=0.27]{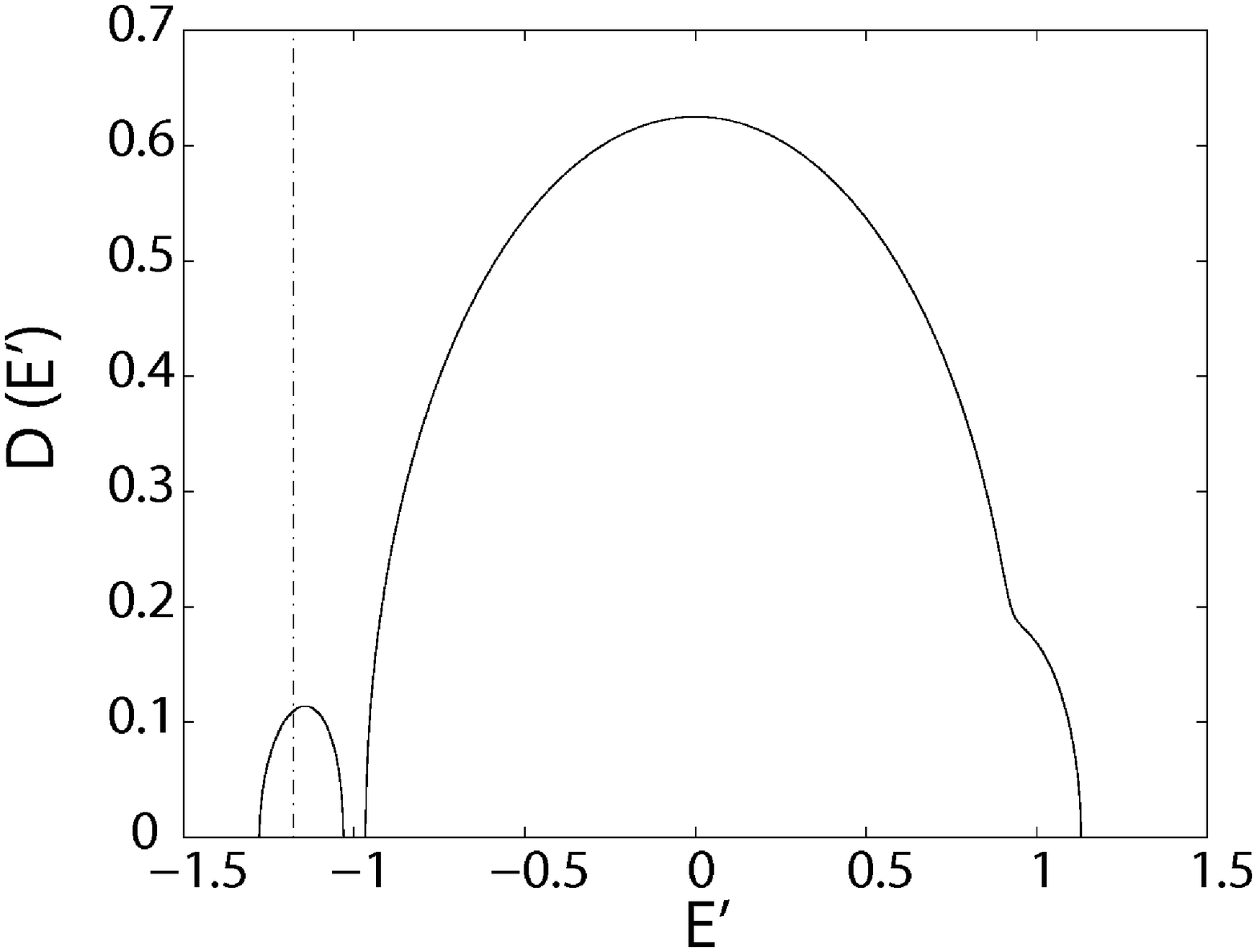}}
    \end{center}
      \caption{Density of states (DOS) for different temperatures, (a) $k_BT = 0$, (b) $k_BT = 2.5\text{x}10^{-3}W$,
       (c) $k_BT = 4.5\text{x}10^{-3}W$ and (d) $k_BT = 5.4\text{x}10^{-3}W$, when $x = 0.053$, $E_{NM} = E_M = 0$,
        $\mu_B B = 0$, $J_{pd} = 0.5W$  and $\text{p} = 0.3x$.  The solid line curve is the DOS for the spin down
         carrier, the dashed one is the DOS for the spin up and the vertical line shows the Fermi level.
         One can see that the density of states is, relative to a nonmagnetic system, a strong function of temperature.}
  \label{fig:DOS_B0J05x053EM0p03_T}
\end{figure}
The diagrams shows the evolution of the spin dependent density of
states as a function of temperature for four different temperatures
in each figure. The temperature is measured in units of the
bandwidth W. In Fig.~\ref{fig:DOS_B0J04x053EM0p08_T-d} and
\ref{fig:DOS_B0J05x053EM0p03_T-d} one can no longer see the
spin-splitting. The position of the Fermi level is also shown. Note
that all energy parameters are normalized by $W$ ($E' = E/W$). From
Fig.~\ref{fig:DOS_B0J04x053EM0p08_T} and
Fig.~\ref{fig:DOS_B0J05x053EM0p03_T}, one can see that when the
magnetic coupling $J_{pd}$ is very small, the split band disappears
(see Ref.~\onlinecite{Arsenault2} for details at low $J_{pd}$),
whereas in the opposite limit, the spin bands split off and a pseudo
gap-appears.\\
\\
Figure.~\ref{fig:resis_T_x053} shows the resistivity in zero $B$
field as a function of temperature with p as a parameter. Results
for two values of $J_{pd}$ are presented. The results for lower
$J_{pd}$ than $0.35W$ are similar to
Fig.~\ref{fig:resis_T_x053-d}\cite{Arsenault2}.
\begin{figure}
    \begin{center}
        \subfigure[]{\label{fig:resis_T_x053-d}\includegraphics[scale=0.35]{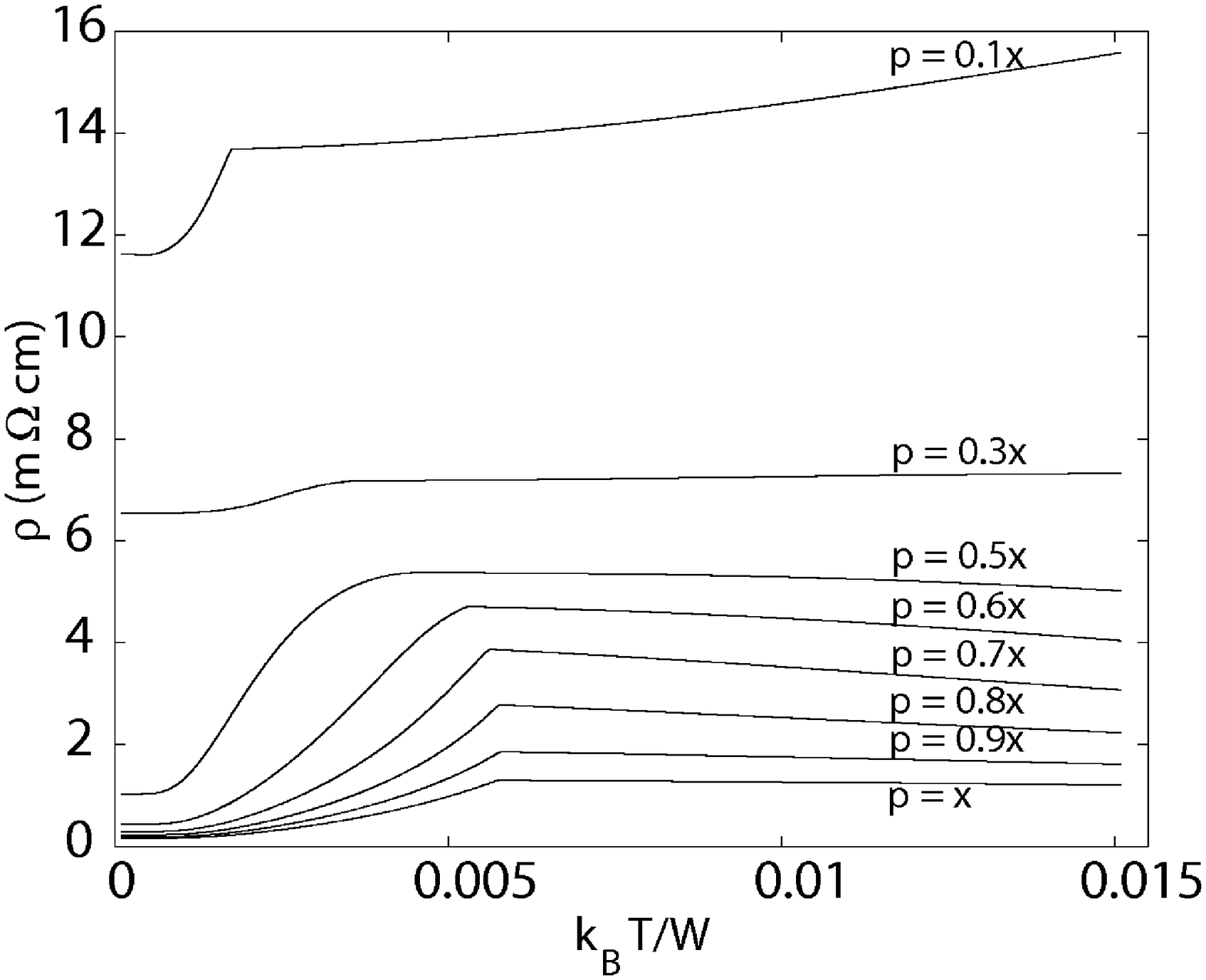}}
        \subfigure[]{\label{fig:resis_T_x053-e}\includegraphics[scale=0.35]{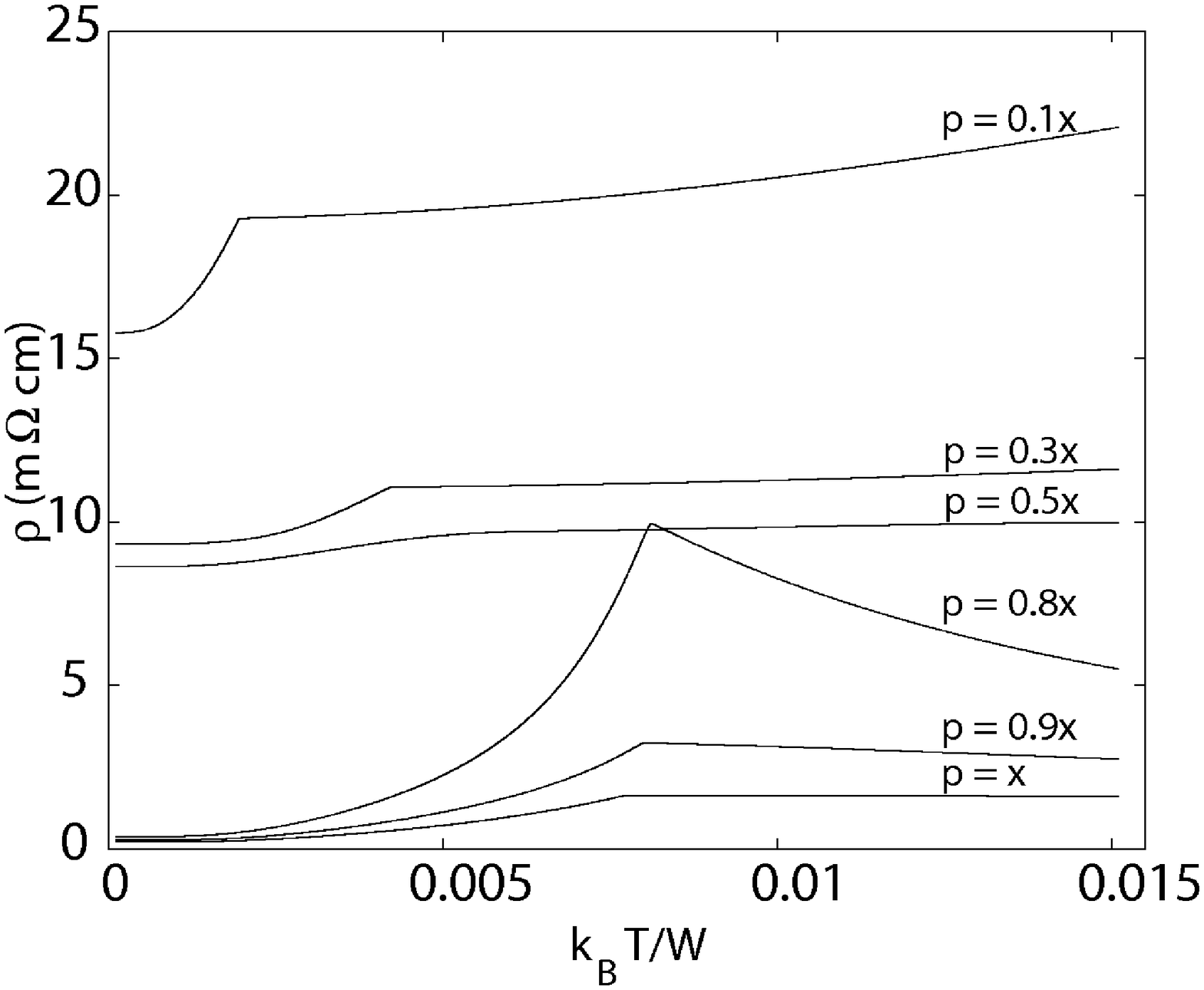}}
    \end{center}
    \caption{Resistivity as a function of temperature for two $J_{pd}$, (a) $J_{pd} = 0.35W$ and (b) $J_{pd} = 0.4W$
      for an impurity concentration $x = 0.053$, $E_{NM} = E_M = 0$ and $\mu_B B = 0$, for two values of $J_{pd}$ ,
      with p as a parameter.}
    \label{fig:resis_T_x053}
\end{figure}
When the impurity bands completely splits from the valence band (
for example for $J_{pd} = 0.5W$ ), the qualitative behavior changes,
as there are now two well defined bands, above and below the Fermi
level (impurity and valence) separated by a gap (see
Fig.~\ref{fig:DOS_B0J05x053EM0p03_T} above and
Ref.~\onlinecite{Arsenault2} ). At high hole concentration
($\text{p}\geq 0.8x$), the system is metallic and resistance
increases with spin disorder, at first rapidly, and then decreases
again above $T_c$. This is intuitively to be expected because at
first, as temperature increases, the disorder increases, and then in
the paramagnetic phase, thermal broadening overcomes the potential
scattering disorder and the lifetime averages out. We should
remember that in the simplest Boltzmann approach the conductivity is
given by
\begin{equation}\label{Bolt_eq}
    \sigma_{xx} = e^2\sum_s\int D_s(E)\left(-\d{\pd f(E)}{\pd E}
    \right)v_s^2(E)\tau_s(E)dE
\end{equation}
and at low temperatures, depending on the product of the scattering
time and the density of states at the Fermi level. In an alloy,
either quantity can change with $B$ and $T$ and determine the
conductivity. Above $T_c$ the spin splitting disappears and the
quantities involved are, in the absence of charged impurity
scattering, only weak functions of temperature. At very low hole
concentration, the Fermi level is in a region of small density of
states where we expect localization. However, CPA is a mean field
method and does not produce localization. Even though the resistance
increases with decreasing density of states at the Fermi level, the
conduction process continues to be band conduction albeit with short
relaxation time. Even if the mean free path reaches the lattice
spacing (random phase limit), the conductivity is still far greater
than hopping conductivity between localized levels\cite{Allen}. The
effect of temperature on resistivity above $T_c$ is not too
significant in the non-localized intermediate cases. The temperature
dependence of the resistance is, in general, a complex interplay of
density of states, velocity and relaxation time (imaginary part of
the self-energy). The temperature induced lowering of the charged
impurity screening length (see Eq.~\eqref{new_self} to
Eq.~\eqref{qsc}) is not included here. In this range of exchange
coupling, $J_{pd}$ does not seem to strongly influence the structure
of the resistivity versus $T$ curves but does change their
magnitude. In Fig.~\ref{fig:resis_T_x053}, for $\text{p} = 0.1x$, a
low carrier concentration, one may see that the resistance increases
with $J_{pd}$. In contrast, in the high hole concentration limit
$\text{p} = x$, there is only a relatively weak variation of
resistance behavior with $J_{pd}$. The complex but regular behavior
of resistance with temperature is a manifestation of
Eq.~\eqref{Bolt_eq}, reflecting the different elements which
determine the value of resistance for a given set of parameters.\\
\\
Figure~\ref{fig:resis_B0EM0_x} shows the resistivity at one temperature, as a function of
impurity concentration for varying degrees of hole doping.
\begin{figure}
    \begin{center}
        \includegraphics[scale=0.35]{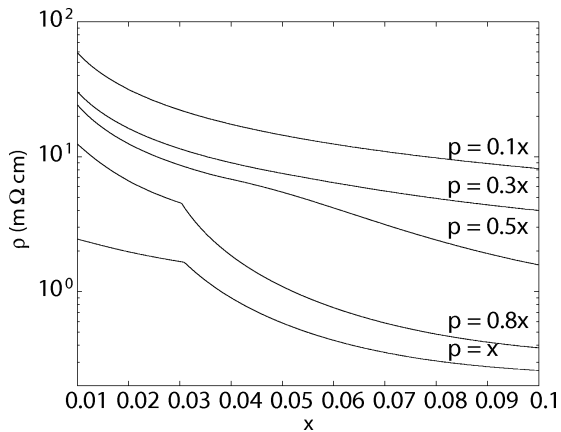}
    \end{center}
    \caption{Resistivity as a function of impurity concentration $x$, for varying degrees of hole doping. The others parameters are $J_{pd} = 0.35W$, $E_{NM} = E_M = 0$, $k_BT = 3.5\text{x}10^{-3}W$ and $\mu_BB = 0$.}
  \label{fig:resis_B0EM0_x}
\end{figure}
The results of Fig.~\ref{fig:resis_T_x053} and Fig.~\ref{fig:resis_B0EM0_x} are based on the spin scattering
model with no additional potential scattering terms, and no
electron-phonon interaction. A fit to experimental data must take
these other mechanisms into account as well. Thus, the complete CPA
self-energy should include other sources of random potential
scattering (nonzero values of $E_{NM}$, $E_M$), in particular, the charged
impurity scattering terms discussed below. We should also add, when
necessary, the electron-phonon self-energy $\Sigma_{ep}(E)$. The imaginary part of
the charged impurity self-energy will contribute another source of
temperature dependent lifetime broadening, and the real part will
enter the density of states.\\
\\
Figure~\ref{fig:resis_B0x053EM0_B_T} shows the resistivity as a function of temperature for
various magnetic fields, for one value of $x$, p, and $J_{pd}$, and again
with $E_M = E_{NM} = 0$.
\begin{figure}
    \begin{center}
        \includegraphics[scale=0.35]{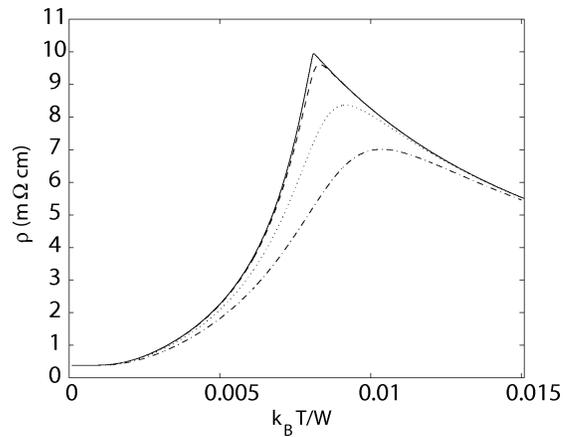}
    \end{center}
    \caption{Resistivity as a function of temperature for various magnetic fields. Others parameters are $x = 0.053$, $E_{NM} = E_M = 0$, $J_{pd} = 0.4W$ and $\text{p} = 0.8x$. The solid line is the zero field result, the dashed line is for $\mu_BB = 1\text{x}10^{-5}W$, the dotted line is for $\mu_BB = 1\text{x}10^{-4}W$ and the dashed-dotted line is for $\mu_BB = 3\text{x}10^{-4}W$.}
  \label{fig:resis_B0x053EM0_B_T}
\end{figure}
Apart from very low and very high temperature, increasing magnetic
field decreases the resistivity, by favoring the alignment of the
impurity spins. The magnetic field also eliminates the sharp
metal-insulator transition around $T_c$, replacing it with a smooth
transition.\\
\\
Figure~\ref{fig:MR_B_J035x053EM0} shows the relative magnetoresistivity defined by
\begin{equation}\label{MR_def}
    MR \equiv \d{\rho_{xx}(B^{ext}) - \rho_{xx}(B^{ext}=0) }{\rho_{xx}(B^{ext}=0)}
\end{equation}
as a function of $B$, for various value of $T$.
\begin{figure}
        \begin{center}
            \subfigure[]{\label{fig:MR_B_J035x053EM0-a}\includegraphics[scale=0.35]{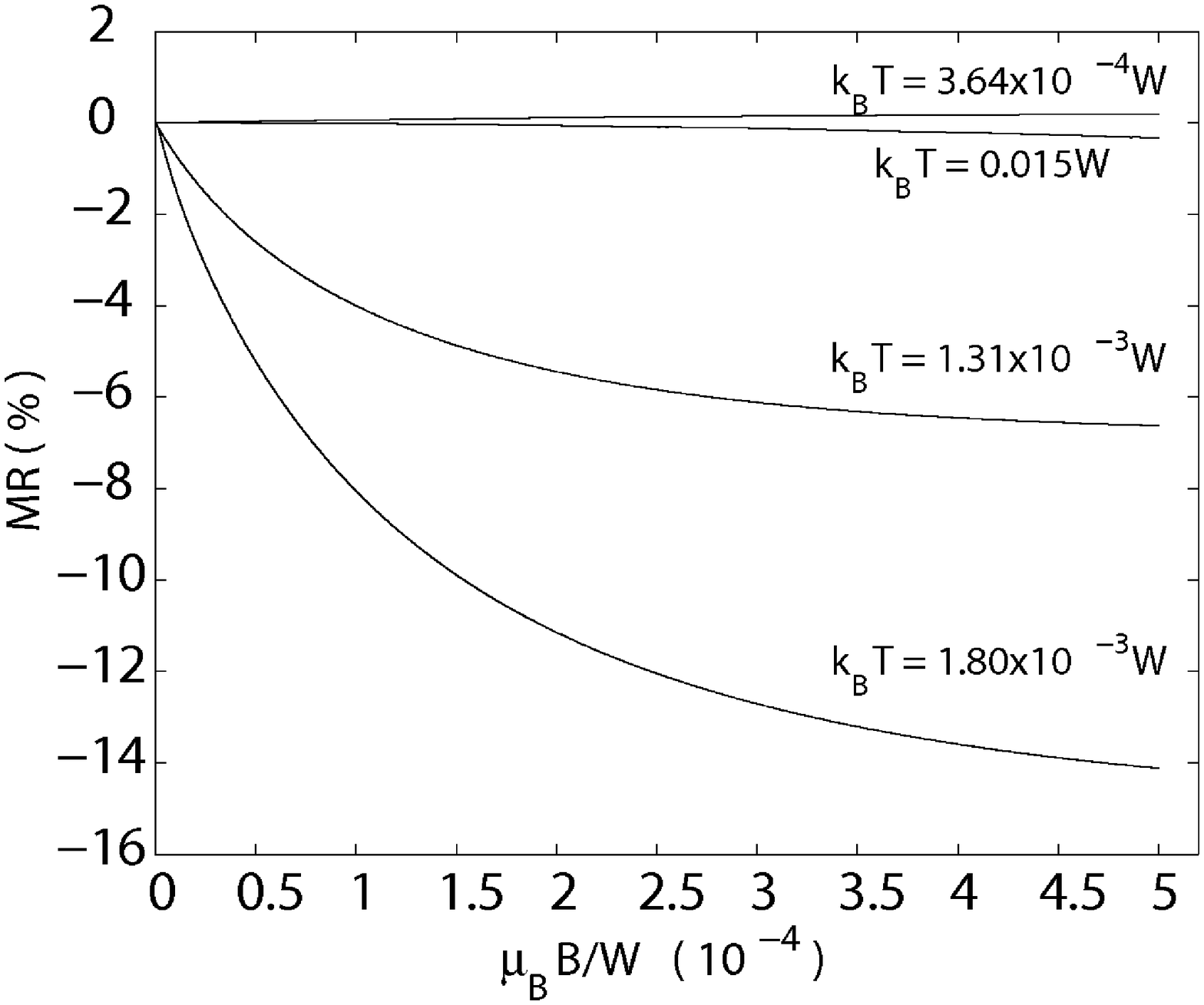}}
            \subfigure[]{\label{fig:MR_B_J035x053EM0-d}\includegraphics[scale=0.35]{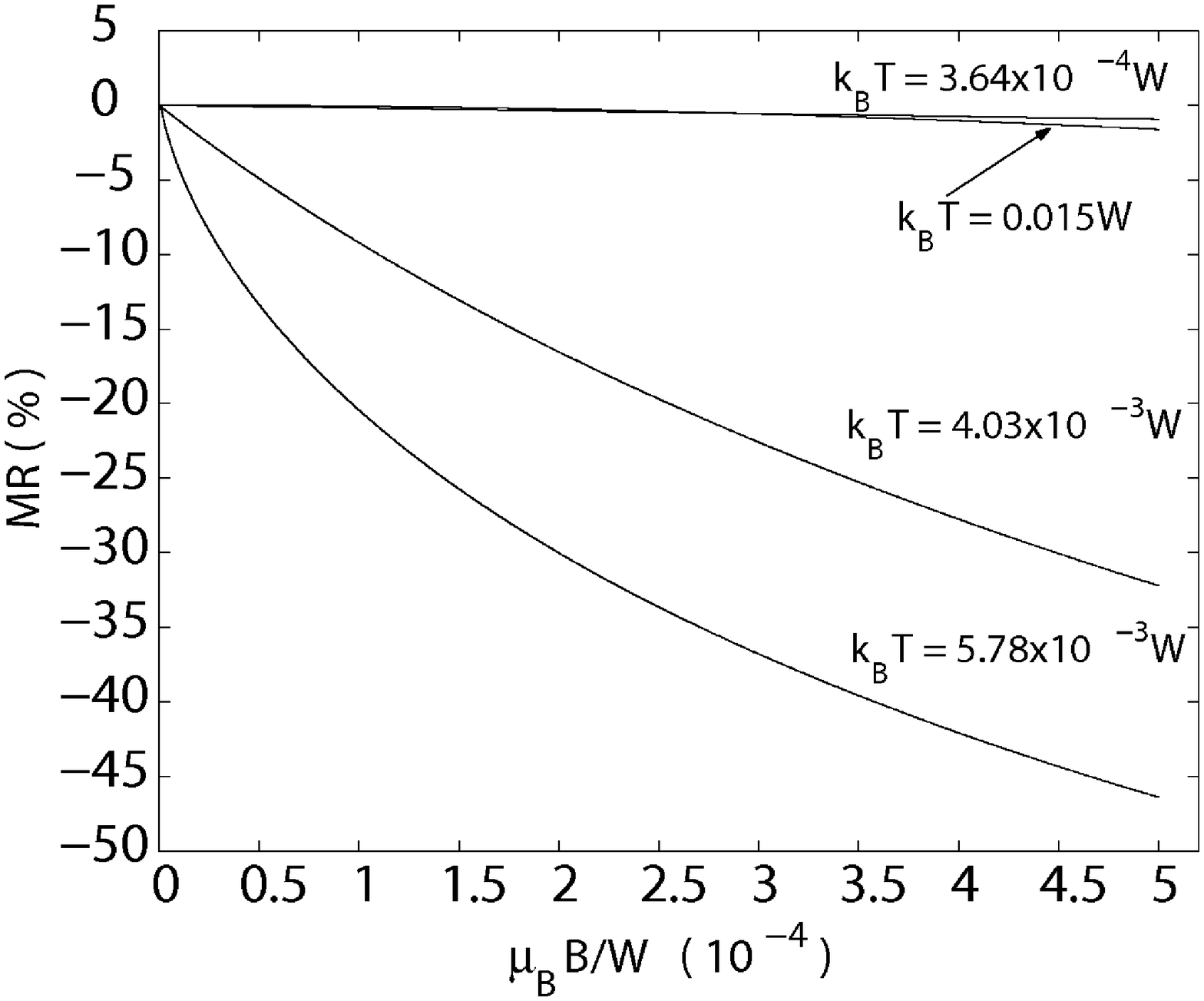}}
        \end{center}
        \caption{Relative magnetoresistivity for two carrier concentration, (a) $\text{p} = 0.1x$ and (b) $\text{p} = 0.8x$,
       for impurity concentration $x = 0.053$, $E_{NM} = E_M = 0$ and $J_{pd} = 0.35W$, with $T$ as a parameter.}
            \label{fig:MR_B_J035x053EM0}
\end{figure}
The overall trend in this exclusively spin scattering model is
strong negative magnetoresistivity, as one would expect, since
increasing Mn spin alignment reduces disorder, and thus reduces the
scattering lifetime, and no other source of scattering are
considered. However, lifetime is not the only quantity entering the
conductivity. As shown in Eq.~\eqref{Bolt_eq}, the density of states $D(E_F)$ at
$E_F$ also plays an important role. There are also regimes of positive
magnetoresistivity. From Fig.~\ref{fig:MR_B_J035x053EM0-a}, we see that it is also possible
for the resistance to increase with $B$ at low $T$ when the hole density
is very low. This is probably because in this limit, the density of
states at the Fermi level decreases with magnetic field and this
effect is stronger than the concomitant increase of the carrier
lifetime due to the suppression of spin disorder with $B$. But we also
know that for low concentrations, when the Fermi level is at the
band edge or in the region of localized states, other changes arise
which are not due to spin disorder scattering and which require
another approach which is based on localization. In the hopping
regime, not describable by CPA, the magnetic field can, for example,
squeeze the localized wavefunctions and increases resistance by
reducing overlap. But it also shifts the energy and the mobility
edge such as to reduce resistance\cite{Movaghar_Roth}. When the
calculated resistance is high and when the Fermi level is in a
region of small density of states $<~ 10^{19}/\text{cm}^3\text{eV}$, we should therefore
not trust the CPA, which gives a good description only of the
metallic regime.
\subsection{Hall effect}\label{subsec:Hall_effect}
By definition, the Hall resistance is given by
\begin{equation}\label{Hall_resis_def}
    \rho_{yx} = \d{-\text{Re}\Big\{\sigma_{yx}\Big\}}{\text{Re}\Big\{\sigma_{xx}\Big\}^2 - \text{Re}\Big\{\sigma_{yx}\Big\}\text{Re}\Big\{\sigma_{xy}\Big\}}.
\end{equation}
Equation~\eqref{Hall_resis_def} being a general definition, the normal and the anomalous spin-orbit generated Hall resistance are included.\\
\\
In the CPA, the sign of the normal Hall effect is not determined by
a simple relationship, but depends on the dispersion of the lattice
and the behavior of the real part of the Green's function at the
Fermi level\cite{Arsenault2,Movaghar_Cochrane1}. A full analytical
analysis of the CPA Hall sign is beyond the scope of this article
but we can state that there is no simple rule. The closest to one is
that $R_H$ is electron-like when the density of states increases
with energy and hole-like when it decreases. In the approximation of
skew scattering, the sign of the anomalous effect follows the sign
of the normal effect as long as the carriers at the Fermi level are
polarized in the same direction as the total magnetization. This is
true here, as can be seen from Fig.~\ref{fig:DOS_B0J04x053EM0p08_T}.
The majority spin band at the Fermi level has the same sign as the
overall magnetization, which is dominated by the localized spins.
Thus, in the phase addition approximation of Eq.~\eqref{phase}, in
order to get the right sign, it is essential to keep the correlation
between spin and energy. The decoupling of the spin magnetization
out of the Kubo formula can give rise to the wrong sign. Here, the
overall polarization, found using Eq.~\eqref{spin_conc}, is opposite
to the direction of the field as it should be. We correct for this
effect by assuming that $\langle\sigma_z\rangle$ is in the same
direction as the overall magnetization.\\
\\
Figure~\ref{fig:rhoH_B_J035x053EM0} shows, for a selected class of
parameters, the overall behavior of the Hall resistivity in the skew
spin scattering model.
\begin{figure}
          \begin{center}
            \subfigure[]{\label{fig:rhoH_B_J035x053EM0-a}\includegraphics[scale=0.35]{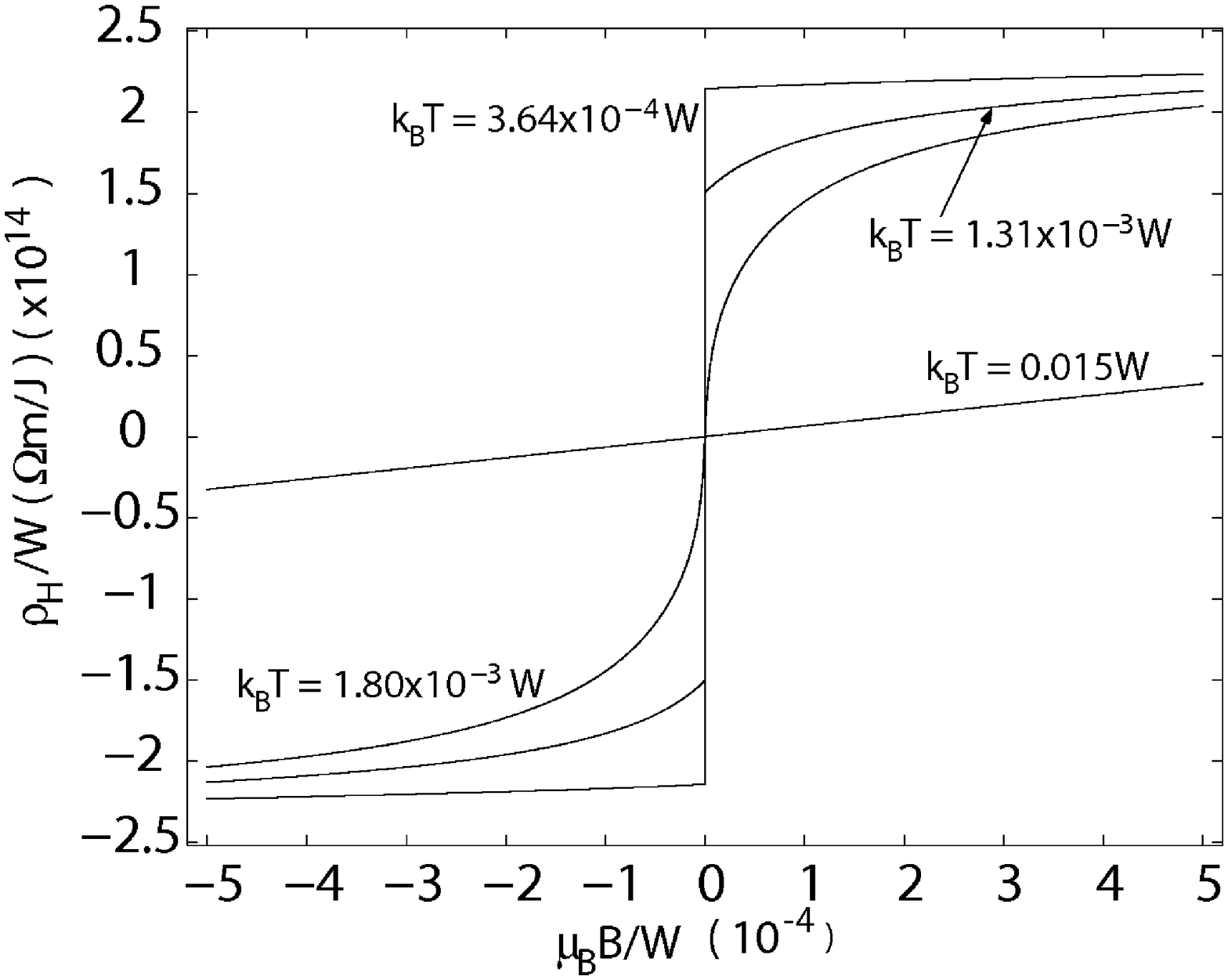}}
            \subfigure[]{\label{fig:rhoH_B_J035x053EM0-d}\includegraphics[scale=0.35]{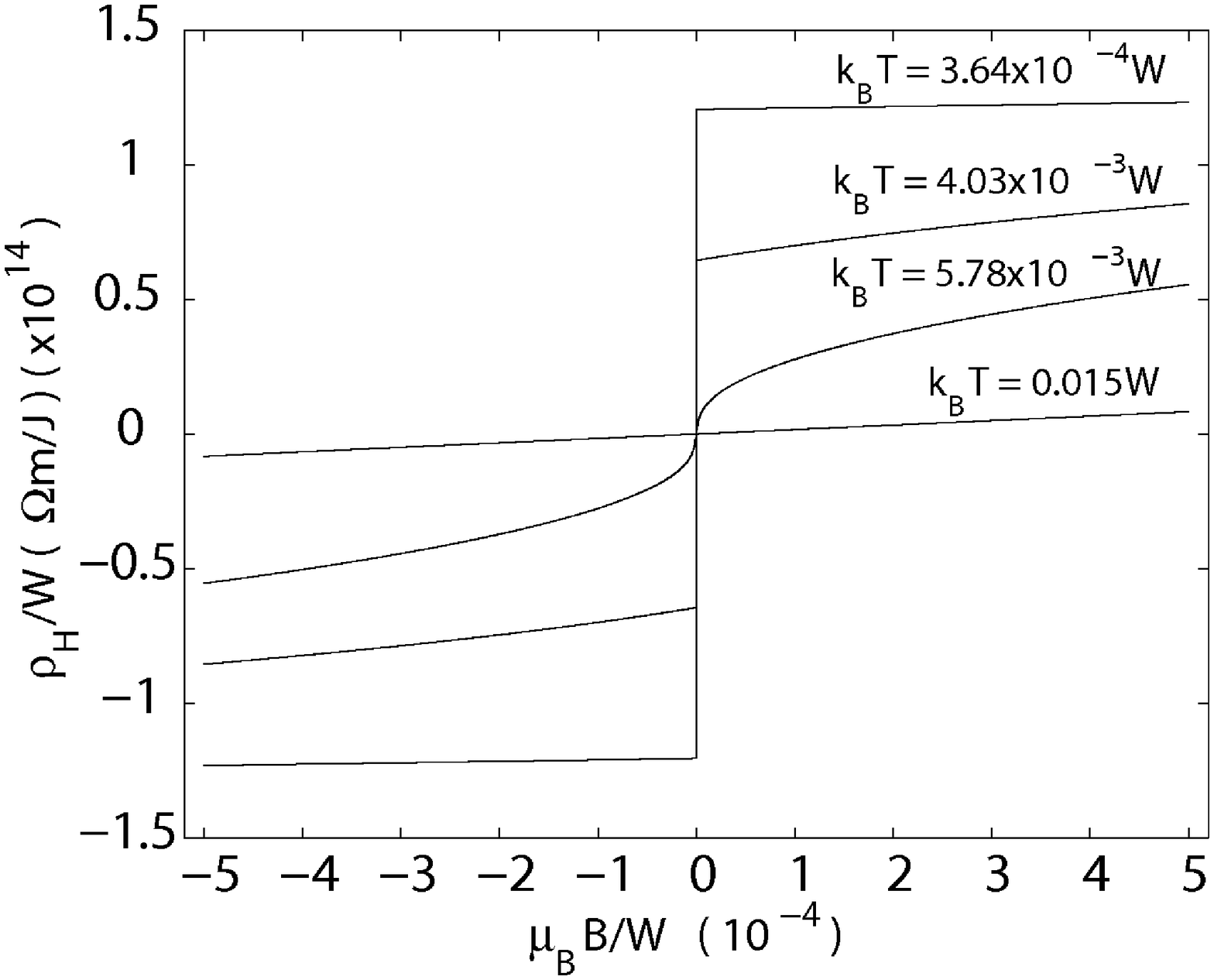}}
          \end{center}
          \caption{Hall resistivity as a function of the applied magnetic field for two carrier concentration, (a)
      $\text{p} = 0.1x$ and (b)  $\text{p} = 0.8x$, for an impurity fraction $x = 0.053$, $E_{NM} = EM = 0$ and
      $J_{pd} = 0.35W$, with $T$ as a parameter}
                \label{fig:rhoH_B_J035x053EM0}
\end{figure}
In order to explain the experimental data we need
$\d{\hbar}{\tau_s}\d{1}{2t} = 0.2$, which means that the skew
scattering energy has to be 0.4 times the tight binding overlap.
This is a very (too) strongly enhanced skew scattering rate. It
suggests that the correct interpretation of the AHE in GaMnAs is
most likely the intrinsic mechanism proposed by Sinova et
al.\cite{Sinova1}. In this work, the AHE is due to the intrinsic
spin-orbit field, and relies on the multiple band nature of this
class of semiconductors. The usual simple nearest neighbor
tight-binding model only gives a skew scattering contribution. The
universality and order of magnitude of the AHE in ferromagnets
suggests that the intrinsic process dominates in most cases.\\
\\
At high temperature, the magnetism disappears. The normal Hall
conduction, linear in $B$ field, is recovered (see the $k_BT =
0.015W$ results Fig.~\ref{fig:rhoH_B_J035x053EM0-a} and
Fig.~\ref{fig:rhoH_B_J035x053EM0-d}).
 Note that for simplicity of notation, we refer to the applied field in the figure as B, even if it was
defined otherwise previously.
\section{Experimental relevance}
Figure~\ref{fig:recapitulatif} shows results of the CPA calculations
for the three transport parameters: resistance, magnetoresistance
and Hall effect in a range of parameters for which a behavior close
to the ones observed experimentally by Ruzmetov\cite{Ruzmetov} and
Ohno \emph{et al}.\cite{Ohno3} is observed. We have added a constant
contribution to the resistivity so that the relative change in
resistivity, such as discussed in connection with
Fig.~\ref{fig:resis_T_x053}, between $T_c$ and $T = 0$ is of the
same order of magnitude as observed in experiment.
\begin{figure}
\begin{center}
\subfigure[]{\label{fig:recapitulatif-a}\includegraphics[scale=0.35]{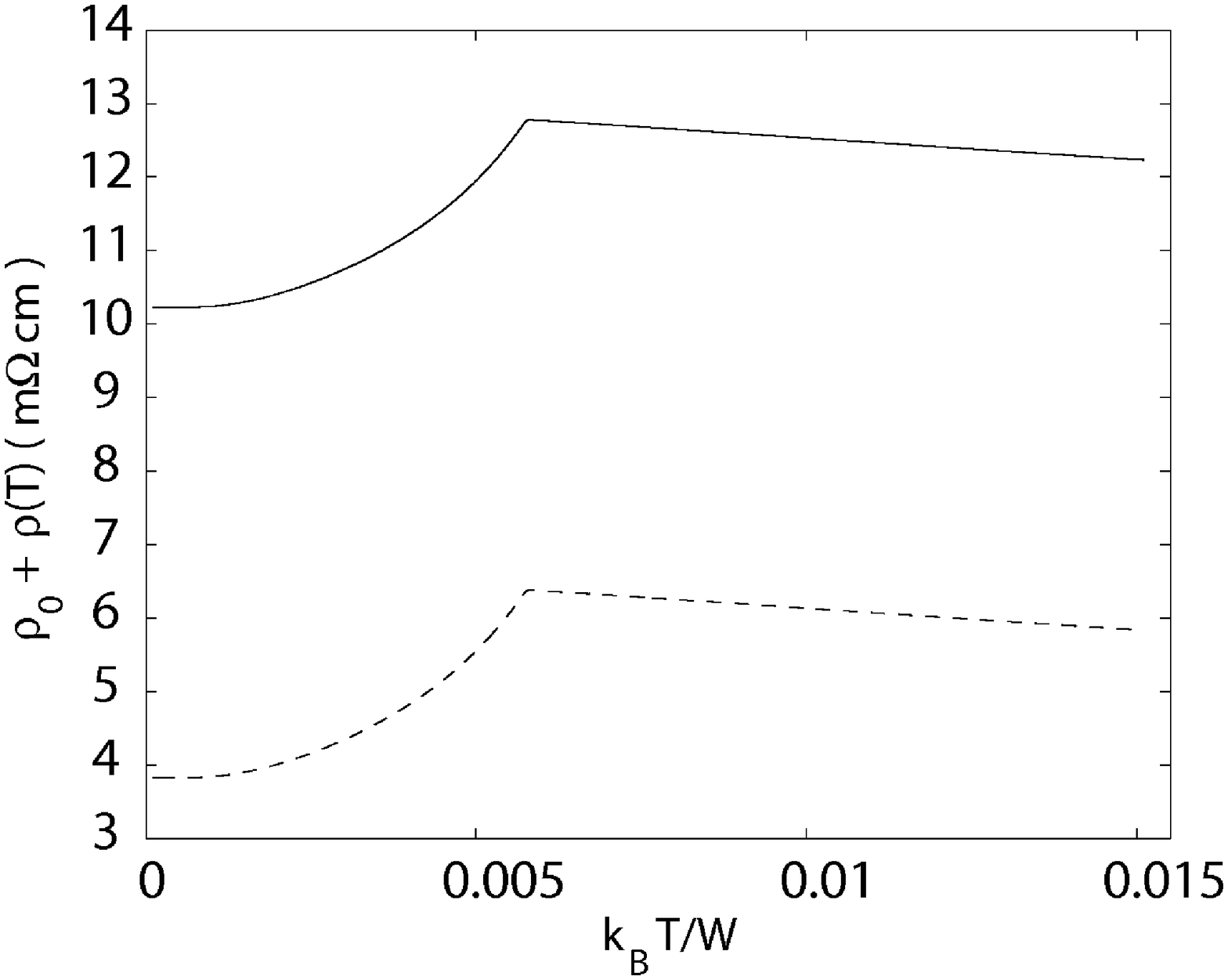}}
\subfigure[]{\label{fig:recapitulatif-b}\includegraphics[scale=0.35]{MLSANS_MR_x053EM0J035p08_B_1_fig6b_fig8b_new.eps}}
\subfigure[]{\label{fig:recapitulatif-c}\includegraphics[scale=0.35]{ML_resis_hall_x053EM0J035p08_B_1_fig7b_fig8c_new.eps}}
\end{center}
\caption{Transport coefficients which can be directly compared to
experiment, (a) $\rho = \rho_0 + \rho (T)$,
       (b) $MR(B)$ and (c) $\rho_H(B)$, for $E_{NM} = E_M = 0$, $J_{pd} = 0.35W$ and $\text{p} =
       0.8x$. Two possible constant contributions are shown in (a), the solid line curve has $\rho_0 = 10$m$ \Omega$cm and the dashed line curve has $\rho_0 = 3.6$m$ \Omega$cm}
\label{fig:recapitulatif}
\end{figure}
The constant term is chosen to be of the same order of magnitude as
the spin scattering rate. The qualitative temperature structure is
satisfactory in the 'metallic regime' of the material, but because
we have neglected the charged impurity scattering, we are
underestimating the temperature drop of the resistivity with
temperature, at higher temperatures. The latter is due to a
reduction in screening length when the density of states at the
Fermi level increases, (see Eq.~\eqref{qsc}).
Figure~\ref{fig:recapitulatif} also illustrates the strong
temperature dependence of the magnetoresistance, which reaches 40 to
50$\%$ at $k_BT \sim 0.0057W$ (this is $\sim 60K$ for $W = 1eV$),
and how it is intimately connected to the magnetic order as observed
experimentally. For an even clearer picture of the connection see
Fig.~\ref{fig:resis_B0x053EM0_B_T}. The reader should note that Van
Esch \emph{et al}.\cite{Van_Esch1} have observed magnetoresistances
of $\sim$ 500\% at $T = 4K$  and 50\% at $T=20K$ in their more
resistive samples with $\rho \sim 1 \Omega$/cm. In this sample the
magnetization dropped by $\sim 100\%$ between $4K$ and $20K$.\\
\\
Hwang and Das Sarma\cite{Hwang_DasSarma} and Lopez-Sancho and
Brey\cite{Lopez-Sancho_Brey} quite rightly pointed out that when
charged impurities are present, they will normally dominate the
scattering rate and the temperature dependence of resistance in a
wide range of parameters. They then decided to use two different
conductivity models for the transport in Mn doped GaAs, one to
explain the temperature dependence of the resistance which
emphasizes charged impurity scattering, and one for the
magnetoresistance which emphasizes spin scattering. We believe this
is not necessary. It is enough to add the configurationally averaged
charged impurity self-energy to the basis band dispersion
$\varepsilon_{\textbf{k}s} \rightarrow\varepsilon_{\textbf{k}s} +
\Sigma_{\textbf{k}s}$, and then to neglect, in the simplest limit,
the real part of the impurity scattering self-energy
$\Sigma_{\textbf{k}s}$. Thus we replace the CPA self-energy,
$\Sigma_s$, which appears in Eq.~\eqref{conduc}, by
\begin{eqnarray}\label{new_self}
    &&\text{Im}\Big\{ \Sigma_s(E) \Big\} + \text{Im}\Big\{
    \Sigma_{\textbf{k}s}(E)
    \Big\} = \text{Im}\Big\{ \Sigma_s(E) \Big\} +\nonumber\\
    &&\eta\int d\textbf{k}'\left( \d{1}{2\pi}
    \right)^3\left|V_{\textbf{k}-\textbf{k}'}
    \right|^2(1-\cos\theta_{\textbf{k}-\textbf{k}'})\delta\Big( E - \varepsilon_{\textbf{k}'s}
    \Big),\nonumber\\
\end{eqnarray}
where $\eta = \d{2\pi}{\hbar}N_i$, with $N_i$ the impurity
concentration, and
\begin{equation}\label{Vkkprime}
    V_{\textbf{k}-\textbf{k}'} = \d{4\pi Z_{imp}e^2}{\varepsilon\varepsilon_0\left| \textbf{k}-\textbf{k}'
    \right|^2}\d{1}{1+\left( q_{sc}/\left| \textbf{k}-\textbf{k}'
    \right| \right)^2}.
\end{equation}
In Eq.~\eqref{Vkkprime}, $Z_{imp}$ is the impurity charge number,
and
\begin{equation}\label{qsc}
    q_{sc}^2 = 4\pi e^2\sum_s\int dED_s(E)\left( -\d{\pd f(E)}{\pd E} \right)
\end{equation}
is the inverse screening length. Since the results of
Section~\ref{App_CPA} show that $D_s(E)$ at the Fermi level change
with temperature and field, the screening length will change as
well. Note that usually, in Boltzmann transport, one does not take
into account the effect of the real part of the self-energy (the
imaginary part gives the scattering) on the actual band structure
because the concentration of dopants is low. Here the concentration
is not low. The real part should in principle, modify the band gap
and effective masses and can be included in our formalism.\\
\\
We agree with Hwang and Das Sarma\cite{Hwang_DasSarma} that the
complete problem in parameter space is enormously complex. In
effect, we should want the impurity scattering to only represent an
additional self-consistent, density of states dependent, lifetime
process. A fully self-consistent Coulomb potential/spin multiband
CPA is far too complex, and of little value.  In addition, the CPA
band theory cannot account for the hopping regime, as pointed out in
the previous section.
\section{The problem of the intrinsic and extrinsic Hall effect}
In the Kane-Luttinger theory of Jungwirth \emph{et
al}.\cite{Jungwirth1}, the Mn dopants only cause a lifetime
broadening and no change to the Kane-Bloch bandstructure. In this
formalism, the intrinsic spin-orbit interaction is Bloch invariant
and only causes band mixing. All disorder is treated only as a
lifetime effect. This is not so in the CPA where the dopant
scattering causes a substantial renormalization of the bandstructure
via the real part of the self-energy and indeed gives an explanation
of the magnetism. The spin-orbit coupling produces new terms in the
Hamiltonian which allow the electrons jumping or tunneling from one
site to the next site to experience the electric field of the
neighboring atoms. The 3-site processes we invoke are not normally
included in the tight-binding modeling of semiconductors. They
represent only small modifications of the bandstructure, much
smaller than the on-site spin-orbit admixture. The anomalous Hall
mechanism in one band tight-binding is basically one of "skew
scattering", except that the jump is always from orbit to orbit, and
the effective magnetic field can only come from another $3^{rd}$
site. We have seen that the one band TB description is inadequate to
describe the AHE in DMS. On the other hand, the multiple band Kane
and/or Kohn-Luttinger\cite{Kohn_Luttinger} approach seems to work
well. It gives the right order of magnitude with an intrinsic AHE.
This suggests that a TB modeling of the AHE must include the
multiple band aspect and at least a second nearest neighbor overlap.
Since, for example, a five percent GaMnAs alloy cannot have a true
Bloch band structure, in the orbital description, it is the short
range many orbital aspect which must give rise to the intrinsic AHE.

\section{Conclusion}
We have presented a CPA theory which could explain the magnetism in
Mn doped semiconductors in the metallic regime and allows one to
calculate the transport parameters. The numerical evaluation of
resistance, magnetoresistance and Hall coefficient, normal and
anomalous, show that the overall trends observed experimentally are
reproduced by the CPA. The power of the method lies in its
simplicity.  All the information is contained in a spin and energy
dependent self-energy $\Sigma_s(E)$. In good approximation, we may
add the effect of the charged impurity and electron-phonon lifetime
corrections to the calculated CPA self-energy. Of the two, the first
is the most important addition because it explains why the
experimental resistance decreases at high temperatures (in general
more strongly than shown in Fig.~\ref{fig:recapitulatif-a}), when
the ferromagnetism has reduced to paramagnetism, and spin-disorder
scattering should have reached its highest value. The "small"
resistance drop at high temperature seen in
Fig.~\ref{fig:recapitulatif-a} is due to CPA bandstructure
renormalization. The CPA and $t$-matrix methods can be extended to
treat magnetic clusters and evaluate effective localized spin-spin
coupling mediated by the
band.\\
\\
We have also demonstrated how to include charged impurity scattering
within the same formalism. The central aim of this paper was to
achieve an understanding of the important mechanism which determines
the conductivity behavior. For this purpose, we have used a one-band
approach which is adequate to understand the conductivity. For
optical properties and the intrinsic AHE, the many band aspects are
essential.\\
\\
The AHE has been modeled as an extrinsic effect in the framework of
skew scattering. We did this using the tight-binding language. The
intrinsic AHE, as invoked by Jungwirth \emph{et
al}.\cite{Jungwirth1}, cannot be derived using a nearest neighbor TB
formalism, and it is not clear how to recover it in the TB
formalism. The problematic of the intrinsic AHE and the TB model is
an interesting one. It should be re-examined in detail because so
far, the nearest neighbor tight-binding methods have proved useful
as bandstructure descriptions of semiconductors.
\begin{acknowledgments}
The authors gratefully acknowledge financial support from the
Natural Sciences and Engineering Research Council of Canada (NSERC)
during this research. P.D. also acknowledges support from the Canada
Research Chair Program. L.-F. A. gratefully acknowledges Martine
Laprise for help with the figures and Pr.~Alain Rochefort for the
use of his computers.
\end{acknowledgments}

\end{document}